# Constructive Axiomatics in Spacetime Physics Part I: Walkthrough to the Ehlers-Pirani-Schild Axiomatisation

Niels Linnemann[*] and James Read[†]

### Abstract

The Ehlers-Pirani-Schild (EPS) constructive axiomatisation of general relativity, published in 1972, purports to build up the kinematical structure of that theory from only axioms which have indubitable empirical content. It is, therefore, of profound significance both to the epistemology and to the metaphysics of spacetime theories. This axiomatisation is, however, self-consciously terse, rendering it difficult to ascertain whether it succeeds. In this article, we provide a pedagogical walkthrough to the EPS axiomatisation, filling relevant conceptual and mathematical gaps and rendering explicit controversial assumptions. There are two companion papers, in which we discuss the significance of constructive approaches to spacetime structure more generally (Part II), and (with Emily Adlam) consider extensions of the EPS axiomatisation towards quantum general relativity based upon quantum mechanical inputs (Part III).

## Contents

[*]niels.linnemann@uni-bremen.de
[†]james.read@philosophy.ox.ac.uk

# 1   Introduction

The 1972 Ehlers-Pirani-Schild (EPS) axiomatization of (the kinematics of) general relativity (GR) is arguably one of the most well-known and highly-regarded rational reconstructions in all of physics.[1] Characteristically, the scheme is both (i) constructively axiomatic in the sense of Reichenbach,[2]—i.e., builds on a basis of empirically supposedly indubitable posits, and (ii) constructivist in the sense of Carnap,[3]—i.e., proceeds semantically in a linear, non-circular fashion. It is, therefore, an ideal case study for the motivations, merits, and conceptual issues involved in both kinds of such rational reconstructions. More specifically, the EPS axiomatisation is of profound significance both to the epistemology of spacetime theories (for the result seems to show that we can recover the spatiotemporal structure of the world from the input empirically-grounded axioms) and the metaphysics of spacetime theories (for one might use the result to argue that spacetime structure is ontologically reduced to the structures invoked in empirically-grounded axioms).

We commit to the study of the EPS scheme in three papers. The current paper provides a detailed walkthrough of the original work by EPS. Such a detailed presentation of the EPS scheme is prompted by recognition that the original presentation, for all its beauty, leaves out important motivational remarks, proofs, and clarifications; at the same time, the overall rational reconstruction can only properly be assessed if all specific steps are clearly laid out and often-ignored ambiguities are rendered explicit. This paper also provides the actual analysis of the status of the EPS scheme *qua* rational reconstruction. The second paper then draws overall lessons on motivation, merits and issues of constructive approaches to spacetime structure and physics more generally, both from the present case study as well as other programs. The third paper (co-authored with Emily Adlam) uses insights from both the walkthrough for EPS and the paper on general lessons on constructive/constructivist approaches to set up a quantum version of EPS—i.e., a constructive-constructivist axiomatisation of kinematics for a (low-energy) theory of quantum gravity.

To facilitate comparison of our presentation to that of the original EPS paper, we mainly follow the structure of EPS' original presentation; in particular, sections 2, 3, 4, 5, 6 and 7 of this paper concern the respective sections in the original EPS paper on the background assumptions (preamble), construction of differential topology, conformal structure, projective structure, Weyl structure and finally Lorentzian structure. (We also refer to EPS' figures in their original numbering.)[4] Among other things, implicit assump-tions will be made explicit, the form of constructive and constructivist reasoning clarified, missing proofs and alternative pathways upon modifying certain assumptions illustrated. Section 8 concerns immediate evaluative questions regarding the EPS scheme in light of this walkthrough.

---

[1] For well-known work on rational reconstructions, see e.g. Laudan (1978).

[2] See Reichenbach (1969). We discuss further Reichenbach and other historical matters in Part II.

[3] We discuss further Carnap's notion of 'constructivist' below.

[4] All figures have been re-drawn ourselves.

## 2  Setup

The structures from which EPS begin their constructive axiomatisation of GR are the following:

**Definition** (Set of events). *A point set $M = \{p, q, ...\}$ is called a* set of events.

**Definition** (Light rays and particles). *The elements of $\mathcal{L} = \{L, N, ...\}$ and $\mathcal{P} = \{P, Q, ...\}$, where $L, N, ..., P, Q, ...$ are all subsets of $M$, are respectively called* light rays *and* particles.

Two remarks here are in order. First: $M$ is a bare set, not yet structured (say) as a differentiable manifold. Second: the same holds for $\mathcal{L}$ and $\mathcal{P}$. In a sense then, it is too early to call $\mathcal{L}$ and $\mathcal{P}$ 'light rays' and 'particles'.

The decisive structures to be defined consecutively and determined empirically in setting up the kinematical structure of GR as sketched above can be loosely characterised in the terminology of EPS as follows (Ehlers et al., 2012, p. 65):

1. *Differential (topological) structure*: Allowing for the definition of (smoothly) differentiable structures (such as functions) on $M$.[5]

2. *Conformal structure:* A field of infinitesimal null cones defined over $M$.

3. *Projective structure:* A family of curves, called *geodesics*, whose members behave in infinitesimal neighbourhood of each point $p \in M$ as straight lines in projective four-space.

4. *Affine structure:* Projective structure, with the addition that the preferred curves carry preferred *affine parameters* such that, infinitesimally, there is an affine geometry around each point $p \in M$.

5. *Metric structure:* Assigns to any pair of adjacent points of $M$ a number called its *separation*.

The characterisations of the above structures (2)-(4) are of infinitesimal kind, in contrast with the more familiar definitions of these structures in an algebraic fashion. Notably, the former kind of definitions are more suited to the constructivist project, as on the infinitesimal versions the weaker structures do not have to be defined under recourse to stronger structures (which may easily run counter the principle of Carnapian semantic linearity, discussed further below), which, however, is often the case for the more familiar algebraic versions. We will now explicitly show the equivalence of the infinitesimal and the algebraic notions.

---

[5]Although by 'differential structure' EPS mean that which allows for the definition of *smooth* structures (e.g. functions) on $M$, in principle one can countenance weaker notions of differential structure, allowing for the definition of only $C^k$ (for finite $k$) differentiable structures (e.g. functions) on $M$. By buying into EPS' restriction here, one has already imposed certain constraints on the resulting kinematical structures derivable from the ensuing axiomatisation (for example, low regularity versions of general relativity would be excluded).

First, a conformal structure $\mathcal{C}$ is usually defined in the algebraic fashion by presupposing a collection of Lorentzian metrics $g, g', \ldots$ on a differentiable manifold $M$ which satisfy (Matveev and Scholz, 2020, §2)

$$g \overset{\mathfrak{c}}{\cong} g' \quad \Leftrightarrow \quad \exists \text{ a function } f \text{ on } M \text{ s.t. } g' = e^f g.$$

The equivalence between this algebraic definition of conformal structure and the infinitesimal version used by EPS can be seen as follows. A cone around the origin in tangent space is given by $-T^2/a + X^2/b + Y^2/c + Z^2/d = 0$ where we take $\{T, X, Y, Z\}$ to be coordinates in the tangent space $T_p M$ spanned by $\partial/\partial t, \partial/\partial x, \partial/\partial y, \partial/\partial z$ with $\{t, x, y, z\}$ a local coordinate chart on $M$. Now, the cone equation can be rewritten directly as $g_{\mu\nu} X^\mu X^\nu = 0$ with tangent vector $X^0 = T, X^1 = X, \ldots$ where $g_{\mu\nu}$ is a Lorentzian metric, uniquely determined up to a factor (a positive factor if the signature is supposed to stay fixed). At the same time, via the exponential map such a cone structure in tangent space could be taken to define an infinitesimal cone in a small neighbourhood around $p$. This is arguably how an infinitesimal cone structure at a point is associated to one and only one class of conformal metric structure at a point. Problematically though, the exponential map requires an affine connection, which will not be available prior to the construction of conformal structure in the EPS scheme. We thus propose on their behalf to take claims of constructing cone structure by EPS in the following to be also implicit claims about the existence of a suitable isomorphic map between manifold neighbourhood and tangent space (as without the latter, no notion of cone structure can be defined relative to the base manifold to begin with). Even if this posit implied—it does not to our knowledge—that there had to be some notion of exponential map and, with it, of affine connection associated to conformal structure, it can be left empirically unspecified by the conformal structure at this stage which one affine connection and *a fortiori* which exponential map exactly.

Second, an affine connection in the usual algebraic sense engenders a notion of parallel transport, including the notion of a geodesic (a curve along which its own tangent is parallel transported). A converse approach is also possible: given a notion of parallel transport $P[\gamma]_s^t : T_{\gamma(s)} M \to T_{\gamma(t)} M$ with $P(\gamma)_s^s = \mathrm{Id}$, $P(\gamma)_u^t \circ P(\gamma)_s^u = P(\gamma)_s^t$ and a smooth dependence of $P$ on the curve $\gamma$ and points $s, t$, an affine connection can be defined in terms of parallel propagation via the relation[6]

$$\nabla_X V = \lim_{h \to 0} \frac{P(\gamma)_h^0 V_{\gamma(h)} - V_{\gamma(0)}}{h} = \frac{d}{dt} P(\gamma)_t^0 V_{\gamma(t)}|_{t=0}.$$

Third, a projective structure $\mathcal{P}$ is usually defined in the algebraic fashion by piggybacking on a class of affine connections $\Gamma, \Gamma', \ldots$ on the manifold $M$ as follows:

$$\Gamma \overset{\mathfrak{p}}{\cong} \Gamma' \quad \Leftrightarrow \quad \exists \text{ a 1-form } \psi \text{ on } M \text{ s.t. } \Gamma'^\mu{}_{\nu\rho} = \Gamma^\mu{}_{\nu\rho} + \delta^\mu{}_\nu \psi_\rho + \delta^\mu{}_\rho \psi_\nu.$$

The equivalence between this algebraic definition of projective structure and the infinitesimal definition utilised by EPS follows from the facts that

---

[6]See e.g. Knebelman (1951) for details.

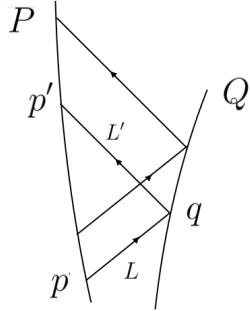

Figure 4 in Ehlers et al. (2012) (own drawing)

(i) two torsion-free affine connections agree on the same geodesics as unparameterised curves iff

$$\exists \text{ a 1-form } \psi \text{ on } M \text{ s.t. } \Gamma'^{\mu}_{\ \nu\rho} = \Gamma^{\mu}_{\ \nu\rho} + \delta^{\mu}_{\ \nu}\psi_{\rho} + \delta^{\mu}_{\ \rho}\psi_{\nu}$$

(see Eastwood (2008), proposition 2.1), and (ii) for any affine connection there can always be found a unique torsion-free affine connection that has the same geodesic structure. (In fact, as we will see later, EPS simply rule out torsionful affine structure by fiat!)

Finally, EPS' definition of metric structure is simply an informal statement of the claim that a metric should be understood as a map $T_p M \times T_p M \to \mathbb{R}$.

## 3    Construction of the differential topology

The first order of business for EPS is to construct differential-topological structure on $M$. As a starting point, EPS make the following assumption:[7]

> We accept in accordance with Axiom $L_1$ that there are 'figures' in $M$ of the type shown in figure 4; i.e., a light signal $L$ emitted from a particle $P$ at $p$ towards another particle $Q$, where it is reflected at $q$ and arrives back on $P$ at $p'$. (Ehlers et al., 2012, p. 70)

Axiom $L_1$ will be discussed below. What is important for us here is the empirically-motivated posit that one can bounce a light ray from the worldline of one particle to that of another and back again. This is arguably the first (idealised) experimental statement in terms of $(M, \mathcal{P}, \mathcal{L})$ that is used to motivate structure on top of the triple $(M, \mathcal{P}, \mathcal{L})$, namely the notions of 'echo' and 'message':

**Definition** (Echo). *The map $e_Q : P \to P, p \mapsto e_Q(p)$ is called an* echo *on $P$ from $Q$.*

---

[7]Note that for better comparability with the original EPS paper, we use the same figure numbers as those used in that article.

**Definition** (Message). *The map $m : P \to Q$, $p \mapsto m(p) = q$ is called a* message *from $P$ to $Q$.*

In their next step, EPS take experimental observations relative to $(M, \mathcal{P}, \mathcal{L})$, *and* the notion of echos to justify the following assumption about the nature of particles $\mathcal{P}$ as well as echos between them:

**Axiom** ($D_1$). *Every particle is a smooth, one-dimensional manifold; for any pair $P$, $Q$ of particles, any echo on $P$ from $Q$ is smooth and smoothly invertible.*

Here, $p$, $q$ and $p'$ are as defined in figure 4.

At this point, it is instructive to clarify what is meant by a *constructivist*[8] approach, and how such an approach proceeds methodologically. A suitable definition of a constructivist system—perfectly compatible with the ambitions of EPS—is given in (Carnap, 1967, §2) (note that Carnap uses the word 'constructional' where we use 'constructivist'):

> To reduce [notion] $a$ to [notions] $b, c$ or to construct $a$ out of $b, c$ means to produce a general rule that indicates for each individual case how a statement about $a$ must be transformed in order to yield a statement about $b, c$. This rule of translation we call a construction rule or constructional definition (it has the form of a definition; cf. §38). By a constructional system we mean a step-by-step ordering of objects in such a way that the objects of each level are constructed from those of the lower levels. Because of the transitivity of reducibility, all objects of the constructional system are thus indirectly constructed from objects of the first level. These *basic objects* form the *basis* of the system.

The basic objects on the EPS account can then be identified as given by a set of events, while novel structure is defined in one of the following three ways ("construction takes place through definition" as Carnap (1967) would say):

1. Through 'explicit definitions' that are of mere abbreviational character (such as that of 'echo' just given),[9]

2. Through 'definitions in use' that define a notion through how it is to feature in propositions (the definition of affine structure below is an example of this), and

3. Through brute ascriptions of specific mathematical properties to and between structures already defined (the enrichment of the notion of echo by means of axiom $D_1$ is an example of this form)—accounting for (idealised) empirical observations.

Axioms of EPS are essentially instances of (3) but may also involve novel structure in terms of (1) or (2).

With all of this in mind, we proceed now to the next EPS axiom:

**Axiom** ($D_2$). *Any message from a particle $P$ to another particle $Q$ is smooth.*

Note that $D_2$ links up the manifold structures that have been associated to individual particles through $D_1$.

---

[8] Recall the distinction between 'constructivist' and 'constructive' presented in the introduction.
[9] We are here and in the next point following the terminology of (Carnap, 1967, §§39-40).

**Definition** (Radar coordinates). *Radar coordinates between two points $P$ and $P'$, denoted as $x_{PP'}$, are charts on $M$ for subset $U$ such that a point $p \in U$ is coordinatised as $(u, v, u', v')$ where $u$ and $v$ are respectively the emission and arrival times at $P$ and $u'$ and $v'$ respectively the emission and arrival times at $P'$. A point on $P$ specifically is coordinatised as $(u, u, u', u')$ (since the emission and arrival times on $P$ are identical).*

The definition is well-defined: $u'$ and $v'$ are determined uniquely by $u$: starting from $u$, one can determine $u'$ and $v'$ uniquely from $u$ since there is—again anticipating axiom $L_1$—one unique line connecting the event at $u$ on $P$ and the event $u'$ on $P'$, and one unique line connecting the event at $u$ on $P$ and the event $v'$ on $P'$.

The third axiom is as follows:

**Axiom** ($D_3$). *There exists a collection of triplets $(U, P, P')$ where $U \subset M$ and $P, P' \in \mathcal{P}$ such that the system of maps $\{x_{PP'}|_U\}$ is a smooth atlas for $M$. Every other map $x_{QQ'}$ is smoothly related to the local coordinate systems of that atlas.*

Several remarks on axiom $D_3$ are in order:

1. Axiom $D_3$ is meant to turn $M$ into a smooth manifold. From now on, $M_p$ denotes the tangent (vector) space of $M$ at $p$. $D_p$ denotes the projective three-space canonically associated with $M_p$. However, a unique manifold is thus only determined if it is furthermore assumed that it satisfies the Hausdorff condition (see Lee et al. (2009))—satisfaction of the Hausdorff condition is thus an implicit further assumption made by EPS![10]

2. According to EPS, axiom $D_3$ is a 'theorem' for Minkowski spacetime. The so-called theorem, however, follows trivially from the fact that Minkowski spacetime is (also) a manifold structure. The assumption of $D_3$ for the general case thus demonstrates a sense in which some of the derived structure in Minkowski spacetime is still imposed onto the more general spacetime structure by EPS—namely, $M$ should still be (at least) a smooth manifold.

3. The second sentence in $D_3$ follows from there being an atlas; it thus only seems to be there in the service of clarity.

4. The coordinates of each $e \in M$ invoked in axiom $D_3$ are assigned as per figure 5. Note that this invokes the echo-and-message construction of figure 4.

Given the definitions and axioms presented thus far, EPS then make the following claim. Although they do not prove the claim explicitly, we provide a proof below.

**Claim.** *Every particle is a smooth curve in $M$.*

*Proof.* We divide this proof into numbered steps, for clarity:

---

[10]Arguably, purely from what one is empirically entitled to say at this stage it seems in fact required to discard the Hausdorff condition—this, however, is not what EPS do. In order to avoid deviating too significantly from EPS' construction, we also simply assume satisfaction of the Hausdorff condition. To see what GR would look like without this condition, see Luc (2020); Luc and Placek (2020)).

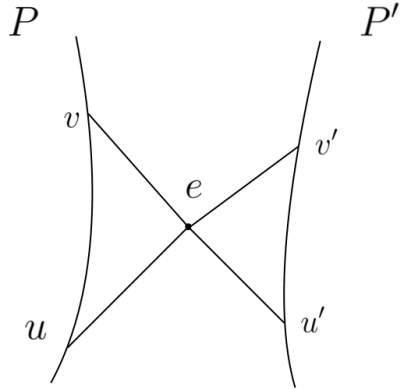

Figure 5 in Ehlers et al. (2012) (own drawing)

1. *Claim:* Every particle $P$ (a one-dimensional manifold) is also a one-dimensional smooth submanifold $M'$ of $M$ (with an atlas borrowed from $M$).

   *Proof:* A smooth submanifold is defined as follows:

   A subset $S \subset M$ is a $k$–dimensional *smooth submanifold* of $M$ if and only if for every $p \in S$, there is a chart $(\phi, U, V)$ around $p$ of $M$ such that

   $$\phi(U \cap S) = V \cap (\mathbb{R}^k \times \{0\}) = \{x \in \phi(U) | x^{k+1} = .... = x^n = 0\},$$

   where $U \subset M$, $V \subset \mathbb{R}^n$ and $n = \dim(M)$. (Wang 2020)

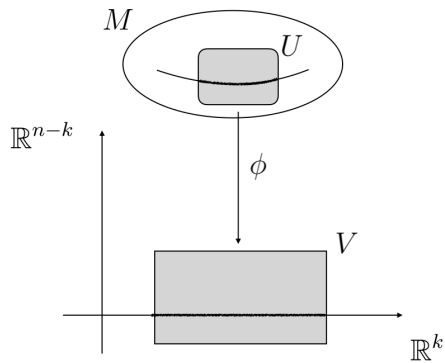

Figure: Smooth submanifold definition (own drawing based on Wang (2020))

$x_{PP'}$ defines charts on $M$ relative to two particles $P$ and $P'$ for some neighbourhood $U$; a point $p \in U$ is coordinatised as $(u, v, u', v')$ where $u$ and $v$ are respectively the emission and arrival times at $P$, and $u'$ and $v'$ respectively the emission and arrival times at $P'$. A point on $P$ specifically is coordinatised as $(u, u, u', v')$ (since the emission and arrival times on $P$ are identical). Note that $u'$ and $v'$ are uniquely determined by $u$: starting from $u$, one can uniquely determine $u'$ and $v'$ through the unique line connecting the event at $u$ on $P$ and the events $u', v'$ on $P'$ respectively. We can thus switch to another coordinate system which has all but one entry equal to zero on $P$. But this means that for every $p \in P$ there is a chart $(\phi, U, V)$ around $p$ of $M$—namely, the one just constructed—such that

$$\phi(U \cap M') = V \cap (\mathbb{R}^1 \times \{0\}) = \{x \in \phi(U) | x^2 = .... = x^4 = 0\},$$

where $U \subset M$ and $V \subset M'$. Thus, every particle $P$ (a one-dimensional manifold) is also a one-dimensional smooth submanifold $M'$ of $M$.

2. Second, $P$ is also a manifold in its own right (in fact, this is captured by axiom $D_1$ of EPS, but would follow now also from the fact that $P$ is a submanifold of $M$.). Then, the inclusion of $i : P \to M$ is smooth since $\phi \circ i \circ \psi^{-1} = j$ is smooth (where $\phi$ is a local chart on $M$, $\psi$ is a local chart on $P$ and $j : \mathbb{R}^1 \to \mathbb{R}^4 : x_1 \to (x_1, 0, 0, 0)$).[11]

3. Third, the map $g : \mathbb{R} \to P$ is smooth (axiom $D_1$).

4. The curve $i \circ g : \mathbb{R} \to M$ is smooth as $i$ is smooth (step 2), and $g$ is smooth (step 3). It is in this sense that we can say that a particle is a smooth curve in $M$.

$\square$

This claim established, there is one final EPS axiom of differential topology:

**Axiom** ($D_4$). *Every light ray is a smooth curve in $M$. If $m : p \to q$ is a message from $P$ to $Q$, then the initial direction of $L$ at $p$ depends smoothly on $p$ along $P$.*

There are some important questions which one might raise regarding the empirical status of axiom $D_4$. First: why assert that every light ray is a smooth curve in $M$, when the analogue for particles was derived from more fundamental axioms? The answer to this question is likely the following: in the case of particles, one can use the radar coordinate construction of Figure 5, and the availability of two particles $P$ and $P'$, in order to derive this result. However, this radar coordinate construction is unavailable in the case of light rays, meaning that the assumption must be inserted 'by hand', by way of an axiom. (This is related to our inability to measure the one-way speed of light: just as we have no way of ascertaining the exact path traversed by a light ray in one direction, we likewise have no way of ascertaining whether this path is smooth. There is room, therefore, to argue that axiom $D_4$ is not in the strict spirit of a constructive axiomatic approach. For more on foundational issues regarding measuring the one-way speed of light, see Salmon (1977).)

---

[11] Recall that a function between two manifolds is said to be smooth if and only if it is smooth in coordinates.

The second question to be raised regarding axiom $D_4$—related to the first—is this: why has smoothness of light rays and particle trajectories been required so far, rather than simply $n$-times differentiability? Especially given that the scheme is to track observations, one might think that even a topological manifold simpliciter or even just a discrete lattice structure should be sufficient. It seems fair to say that the EPS scheme makes at this point a questionable idealisation not easily grounded in the empirical data alone. Lifting this assumption would lead to the possibility of the constructive axiomatisation of e.g. low-regularity GR.

## 4  Light propagation and conformal structure

With the differential-topological structure of $M$ established, EPS now proceed to build up conformal structure from further axioms on the propagation of light. They begin with the following:

**Axiom** ($L_1$). *Any event $e$ has a neighbourhood $V$ such that each event $p$ in $V$ can be connected within $V$ to a particle $P$ by at most two light rays. Moreover, given such a neighbourhood and a particle $P$ through $e$, there is another neighbourhood $U \subset V$ such that any event $p$ in $U$ can, in fact, be connected with $P$ within $V$ by precisely two light rays $L_1, L_2$, and these intersect $P$ in two distinct events $e_1, e_2$ if $p \notin P$. If $t$ is a coordinate function on $P \cap V$ with $t(e) = 0$, then $g(e) := -t(e_1)t(e_2)$ is a function of class $C^2$ on $U$.*

The geometrical constructions envisaged in axiom $L_1$ are presented in figure 6.

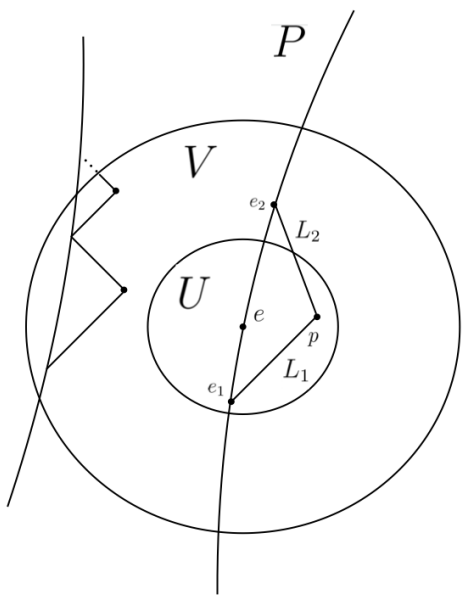

Figure 6 in Ehlers et al. (2012) (own drawing)

There are several remarks which must be made on axiom $L_1$. First, this axiom does not exclude any solution of general relativity, since such a neighbourhood exists locally for all such spacetimes. This is a result due to Perlick:[12]

**Proposition** (Perlick (2008), p. 4). *Let $\gamma$ be a timelike worldine ('clock', in Perlick's terminology) in an arbitrary general relativistic spacetime and let $p = \gamma(t_0)$ be some point on $\gamma$. Then there are open subsets $U$ and $V$ of the spacetime with $p \in U \subset V$ such that every point $q$ in $U \setminus \text{image}(\gamma)$ can be connected to the worldline of $\gamma$ by precisely one future-pointing and precisely one past-pointing light ray that stays within $V$. In this case, $U$ is called a* radar *neighborhood of $p$ with respect to $\gamma$.*

Thus, EPS are not precluding the possibility of recovering the full solution space of general relativity by imposing axiom $L_1$. This, however, leads into our second remark on axiom $L_1$: Pfister and King (2015) point out that this axiom cannot obtain globally in all solutions of general relativity:

> It should also be stressed here that, in strong gravitational fields, axiom $L_1$ is no longer valid globally: effects like the light deflection by big masses can lead to more than two light connections between $p$ and $P$, and so-called horizons can have the effect that only one light connection exists, or none at all. Axioms $L_1$ and $L_2$ already contain a part of Einstein's equivalence principle, so essential to general relativity: in gravitational fields, light has locally the same behavior as in special relativity. (Pfister and King, 2015, p. 56)

The third remark on axiom $L_1$ is this. The assumption that $g$, as defined in the axiom, is a function of class $C^2$ on $U$ is (again) a strong one. Upon weakening this assumption, the EPS scheme can be liberalised to a constructive scheme for Finslerian spacetimes (of which Lorentzian spacetimes are a special case): see Lämmerzahl and Perlick (2018) and Pfeifer (2019). Consequently, it is implausible that Lorentzian spacetimes rather than Finslerian spacetimes are singled out from mere empirical considerations *à la* EPS.

These remarks made, we turn now to axiom $L_2$:

**Axiom** ($L_2$). *The set $L_e$ of light-directions at an (arbitrary) event $e$ separates $D_e \setminus L_e$ into two connected components. In $M_e$ the set of all non-vanishing vectors that are tangent to light rays consists of two connected components.*

Recall again that $M_e$ denotes the tangent (vector) space of $M$ at $e$ as introduced above and $D_e$ the associated projective three-space. Axiom $L_2$ establishes a purely point-based distinction between vectors, conceptually speaking a point-based predecessor to a distinction between space and time. At least a (locally imposed) conformal connection would be needed as surplus

---

[12]On the proof of the following proposition, Perlick writes:

"To prove this, we just have to recall that every point in a general-relativistic spacetime admits a convex normal neighborhood, i.e., a neighborhood $V$ such that any two points in $V$ can be connected by precisely one geodesic that stays within $V$. Having chosen such a $V$, it is easy to verify that every sufficiently small neighborhood $U$ of $p$ satisfies the desired property." (Perlick, 2008, p. 134) structure to 'continuously' link the thus-established distinction at the tangent space of one point to that at the tangent space of a neighbouring point.

All axioms are now in place to allow for the construction of conformal structure. Schematically, the construction has the following intermediate steps:

1. Explore the mathematical properties of $g$. In particular, construct a rank-two tensor $g_{\mu\nu}$ from $g$.

2. Use observational data to select a preferred non-degenerate quadric associated with $g_{\mu\nu}$.[13] This non-degenerate quadric will express the signature on $M$ as well as provide a threefold distinction of the elements of $\mathcal{D}_e$ (extending the distinction between vectors introduced with axiom $L_2$).

3. Show that this quadric can be expressed in terms of a tensor density.

On (1), EPS begin with a lemma, stated and proved below:

**Lemma** (Properties of $g$). *The following statements hold about $g$:*

*(a) $g(p) = 0$ (for $p \in U$) if and only if $p$ lies on a light ray through $e$.*

*(b) $g_{,\mu}(e) := \partial_\mu g(e) = 0$.*

*(c) $g_{\mu\nu}(e) := g_{,\mu\nu}(e) := \partial_\mu \partial_\nu g(e)$ defines a tensor at $e$.*

*(d) The tangent vector $T^\mu$ of any light ray $L$ through the point $e$ satisfies $g_{\mu\nu} T^\mu T^\nu = 0$.*

*(e) $g_{\mu\nu} \neq 0$ on particles.*

*Proof. ad (a):* Left-to-right: Suppose $g(p) = -t(e_1) t(e_2) = 0$. Thus, either $t(e_1) = 0$ or $t(e_2) = 0$. $t : V \cap P \to \mathbb{R}$ is a coordinate function with $t(e) = 0$; since it is a coordinate function, it has to be strictly monotonic and so cannot be zero at any other event than $e$. Hence, either $e_1 = e$ or $e_2 = e$. So $p$ lies on a light ray through $e$. Right-to-left: $g(p) = -t(e_1) t(e_2)$. If $p$ lies on a light ray through $e$, then $e_1$ is $e$, so $g(p) = -t(e) t(e_2)$. But $t(e) = 0$, so $g(p) = 0$.

*ad (b):* Suppose for contradiction that $g_{,\mu}(e) \neq 0$.[14] Consider the hypersurface picked out by $g(p) = 0$ for points $p$ in the neighbourhood $U$ of $e$. This is in fact a smooth hypersurface around $e$ since the normal $g_{,\mu}(e) \neq 0$ is non-zero (in coordinates).[15] This would contradict the second part of axiom $L_2$, which states that the set of all non-vanishing vectors that are tangent to light rays should consist in two connected components: this cannot be the case if the hypersurface in question is smooth as the tangent space at $e$ is then the standard tangent space of a smooth (sub)manifold, which is in particular connected. Thus, it must be the case that $g_{,\mu}(e) = 0$.

---

[13] A quadric is a $D$-dimensional hypersurface in a $(D + 1)$-dimensional space that is determined by the zero set of an irreducible second degree polynomial in $D + 1$ variables.

[14] In deploying here proof by contradiction, we (like EPS) explicitly do not commit to a constructive approach to *mathematics*, but only to the axiomatisation of our physical theories, in the sense already articulated above.

[15] The normal to a hypersurface being non-zero is sufficient to render it smooth. This follows from the regular set theorem (see (Tu, 2011, §9.4); the remark under Lemma 9.10 demonstrates nicely that a non-zero normal is not necessary for smoothness). Heuristically, the sufficiency of a non-zero normal for hypersurface smoothness should seem natural from the fact that the implicit function theorem can be used to establish the smoothness of a level set (such as that one defining a hypersurface).

*ad (c):* One can demonstrate that $g_{\mu\nu}(e)$ has tensorial transformation properties—and therefore defines a tensor at $e$—as follows:

$$\begin{aligned}
g_{\mu\nu} := \frac{\partial^2 g}{\partial x^\mu \partial x^\nu} &= \frac{\partial}{\partial x^\mu}\left(\frac{\partial}{\partial x^\nu}g\right) \\
&\mapsto \frac{\partial x^{\mu'}}{\partial x^\mu}\frac{\partial}{\partial x^{\mu'}}\left(\frac{\partial x^{\nu'}}{\partial x^\nu}\frac{\partial}{\partial x^{\nu'}}g\right) \\
&= \frac{\partial x^{\mu'}}{\partial x^\mu}\frac{\partial^2 x^{\nu'}}{\partial x^{\mu'}\partial x^\nu}\frac{\partial}{\partial x^{\nu'}}g + \frac{\partial x^{\mu'}}{\partial x^\mu}\frac{\partial x^{\nu'}}{\partial x^\nu}\frac{\partial}{\partial x^{\mu'}}\frac{\partial}{\partial x^{\nu'}}g \\
&= \frac{\partial x^{\mu'}}{\partial x^\mu}\frac{\partial x^{\nu'}}{\partial x^\nu}g_{\mu'\nu'}.
\end{aligned}$$

The first term vanishes in the above since $\partial_\mu g(e) = 0$. So $g_{\mu\nu}$ has the requisite tensorial transformation properties, and therefore defines a tensor at $e$.

*ad (d):* Differentiating twice with respect to $s$ from $g(x^\mu(s)) = 0$, we have:

$$\frac{\partial g}{\partial x^\mu}\frac{dx^\mu}{ds} = 0$$

$$\Rightarrow \frac{d}{ds}\left[\frac{\partial g}{\partial x^\mu}\frac{dx^\mu}{ds}\right] = 0$$

$$\Rightarrow \frac{\partial^2 g}{\partial x^\mu \partial x^\nu}\frac{dx^\mu}{ds}\frac{dx^\nu}{ds} + \frac{\partial g}{\partial x^\mu}\frac{d^2 x^\mu}{ds^2} = 0$$

$$\Rightarrow g_{\mu\nu}T^\mu T^\nu = 0.$$

where here $T^\mu := \frac{dx^\mu}{ds}$ and again we have used that $\partial_\mu g(e) = 0$.

*ad (e):* If $x^\mu(t)$ is the parameter representation of $P$ in any permissible coordinate system, then $g\{x^\mu(t)\} = -t^2$; hence, $g_{\mu\nu}K^\mu K^\nu = \partial_t^2(-t^2) = -2$, $K^\mu$ being the tangent vector of $P$ at $e$ with respect to $t$ (so effectively $K = \partial/\partial t$). Hence, $g_{\mu\nu} \neq 0$ on particles. □

This lemma established, we turn now to step (2): associating a preferred quadric with $g_{\mu\nu}$. We first prove another lemma:

**Lemma** ($L_e \subseteq Q$). *The set of light rays at event $e$, $L_e$, is contained in the 3-dimensional hypersurface $Q$ defined in $D_e$ by $g_{\mu\nu}T^\mu T^\nu = 0$ (called a quadric).*

*Proof.* This result follows from the fact that light rays on $e$ with tangent $T^\mu$ fulfil $g_{\mu\nu}T^\mu T^\nu = 0$ (see lemma (properties of $g$, part (d))). □

We normalise $g_{\mu\nu}$ (new notation: $\mathscr{g}_{\mu\nu}$) such that $\det(\mathscr{g}_{\mu\nu}) = -1$; this leaves the definition of the quadric form $Q$ unaffected ($\mathscr{g}_{\mu\nu}$ induces the same form as $g_{\mu\nu}$). It can then be shown that:

**Claim.** $\mathscr{g}_{\mu\nu}$ *is a tensor density.*

*Proof.* We have already seen that $g_{\mu\nu}$ defines a tensor. $\mathscr{g}_{\mu\nu}$ is a normalised version of $g_{ab}$, transforming as a tensor up to a multiple of the metric determinant. It is therefore a tensor density, in the sense of (Anderson, 1967, §1.7) (cf. (Bergmann, 1942, ch. 16)). □

Using $g_{\mu\nu}$, we can then distinguish between different classes of vectors as follows:

**Definition** (Vector distinctions). *A vector $T^\mu$ is called ($g$-)null/timelike/spacelike if and only if $g_{\mu\nu}T^\mu T^\nu$ is, respectively, equal to, greater than, or less than zero.*

From this, one can then prove the following claim:[16]

**Claim** (Properties of $Q$ and $L_e \subseteq Q$). *The quadratic form $g_{\mu\nu}\zeta^\mu\zeta^\nu$ associated with the quadric $Q$ is (a) non-degenerate (there is no $X^\mu \neq 0$ such that, for all $Y^\mu$, $g_{\mu\nu}X^\mu Y^\nu = 0$), and (b) normal hyperbolic (i.e., of signature $(+,+,+,-)$). Furthermore, (c) the set of light directions is equal to the quadric, i.e. $L_e = Q$.*

*Proof. ad (a)*: Suppose for contradiction that the quadric form is degenerate. Given that $g_{\mu\nu}$ is a symmetric bilinear form by construction, this is equivalent to $g_{\mu\nu}$ being singular. In particular, in some coordinate system $\{e_i\}$ ($i = 1, \dots, n$, where $n$ is the dimension of the manifold), $g_{\mu\nu}$ can be diagonalised with at least one zero on the diagonal. This, however, means that relative to this coordinate system, we can write the quadric equation without at least one of the coordinates. Thereby, the level set is revealed to be of codimension lower than 1, and in particular not a three-dimensional hypersurface (if one is working, as EPS are, in the case $n = 4$) that induces a distinction in directional space for $L_e$ as required by axiom $L_2$ and thus required for $Q$ to capture consistently the structure of light propagation.

*ad (b)*: Assuming non-degeneracy, (Synge, 1965, pp. 17-18) demonstrates that only the $(+,+,+,-)$ quadric signature can separate future and past into two connected components. 'Normal hyperbolicity' just means that the signature of the quadric is $(+,+,+,-)$.[17]

*ad (c)*: No proper subset of any quadric has the required topological properties of separating future and past into two connected components, so $L_e$ must coincide with the entire quadric, yielding $L_e = Q$. To see this, assume that $L_e$ is a proper subset of $Q$ while still a quadric. We first note that $Q$ is defined in terms of a quadric equation in such a way that if a point on the quadric picked out by a corresponding vector $X$ is taken out of the set, then—if the subset is supposed to be a quadric as well—any point in the direction of $X$, i.e. any point picked out by a multiple of $X$, will have to be removed as well. But removing even a single point from $Q$ means that $L_e$ will not allow for splitting the space-like from the time-like directions anymore—in clash with the requirement on $L_e$ from the first part of axiom $L_2$. $\square$

After this setup, EPS can take the resulting $g_{\mu\nu}$ to define a conformal structure (denoted by $\mathcal{C}$ from now on): $g_{\mu\nu}$ stands for an equivalence class of $g_{\mu\nu}$ (see the construction of $g_{\mu\nu}$ above), which itself defines a conformal structure $\mathcal{C}$ in the algebraic sense introduced in section 2.

---

[16]Note that in EPS the definition of the tensor density follows after mentioning this claim; we, however, wish the vector distinction—induced from the quadric $Q$/the tensor density $g$—to be already available at this stage in order to show the non-degeneracy of the quadric $Q$.

[17]Note that order and denotion of + as opposed to − is subject to convention; other conventional choices would fix the same quadric.

One defines (note that this is essentially the same definition of ($\mathscr{g}$-)null/timelike/spacelike as given above; we deploy the following definition in order that our notation align with that of EPS):

**Definition** (Vector distinctions (2))**.** *A vector $T^\mu$ is called ($\mathcal{C}$-)null/timelike/spacelike if and only if $\mathscr{g}_{\mu\nu} T^\mu T^\nu$ is, respectively, equal to, greater than, or less than zero.*

Explicitly, $\mathcal{C}$ is set up via an operational procedure for determining the components of $\mathscr{g}_{\mu\nu}$ relative to some coordinate system which proceeds as follows:

**Claim.** *There exists a basic operational procedure to determine values of the conformal structure $\mathcal{C}$ associated to $\mathscr{g}_{\mu\nu}$ relative to a radar coordinate system for a point $e$.*

*Proof.* The components of $\mathscr{g}_{\mu\nu}$ relative to a coordinate system can be ascertained via the following steps:

1. Note first that nine equations in nine unknowns would be sufficient to fix the nine components of $\mathscr{g}_{\mu\nu}$ (given that $\mathscr{g}_{\mu\nu}$ has one less degree of freedom than a metric $g_{\mu\nu}$).

2. Create a light ray for each direction (hereby, use the notion of linear independence provided by a coordinate basis in order to identify distinct directions). There are three directions.

3. Use these three linearly independent light rays to construct nine linearly independent squares which fix $\mathscr{g}_{\mu\nu}$: the light ray in direction $i$ will be made to hit a point $p_i$ sufficiently close to the emission point so that it can still be coordinatised in radar coordinates. Once the radar coordinates for $p_i$ are known, the corresponding directional vector can be expressed in the coordinates dual to the radar coordinates (i.e., $\partial/\partial T$, $\partial/\partial X$, $\partial/\partial Y$, $\partial/\partial Z$ with $R, T, X, Y, Z$ being the radar coordinates).[18]

$\square$

Clearly, light rays are $\mathcal{C}$-null curves (this follows directly from part (d) of lemma (properties of $g$)). We want to show now, though, that light rays are null *geodesics*. The other major goal is to establish that (local) light ray cones are captured by the $\mathcal{C}$-null cone structure.[19] Towards these goals, EPS first need to establish some intermediate results:

**Claim.** $v_e := \left\{ q \in \mathcal{M} : g(q) = 0, q \text{ is a light ray connected to } e \right\}$ *is a smooth hypersurface, except at $e$ itself.*

---

[18]Practically, it needs to be ensured that the light ray gets re-emitted at $p_i$ in the direction of the observer. This could for instance be guaranteed by simply re-emitting into all directions, together with a label on the signals that they derive from signal in direction $i$ (which must then in turn have, of course, been equipped with a label of its direction as well). More concretely, the function of the label could be realised by the frequency of the light ray chosen, and the function of the re-emitter by an antenna tuned only to that frequency.

[19]Remember that conformal structure can be both be characterised algebraically and geometrically, as explained in section 2.

*Proof.* First show that there exists a neighbourhood $W$ of $e$ such that if $e \neq q \in W$ and $g(q) = 0$, then $g_{,\mu}(q) \neq 0$. For this, note that

$$g_{,\mu}(q) - \underbrace{g_{,\mu}(e)}_{\substack{=0 \text{ as shown in part (b) of lemma (Properties of g)}}} = \int_e^q g_{\mu\nu}\{x^\lambda(u)\}T^\mu(u)du.$$

In particular, at $e$, $g_{\mu\nu}T^\nu \neq 0$; the latter function is continuous in $u$. Thus, for sufficiently small departure from $e$ (in $u$ alternatively speaking), $g_{\mu\nu}T^\nu \neq 0$ as well due to continuity. Thus,

$$g_{,\mu}(q) = \int_e^q \underbrace{g_{\mu\nu}T^\nu}_{:= f(u)} du = \int_e^q f(u)\, du = \underbrace{f(c)}_{\neq 0} \underbrace{[u(q) - u(e)]}_{>0} \neq 0.$$

where we used the mean value theorem.

Second, read $g(q) = g(x^\lambda(q)) = g(x^\lambda(q), q)$, and $g_{,\mu}(q) = g_{,\mu}(x^\lambda(q), q)$ (i.e., through their definition in coordinates) where $g, g_{,\mu} : \mathbb{R}^4 \times M \to \mathbb{R}$. Together with the result from the first step of the proof that $g_{,\mu}(q) \neq 0$, the implicit function theorem then tells us that there exists a unique assignment $U \subset M \to \mathbb{R}^4, q \mapsto x^\lambda(q)$ such that $g(x^\lambda(q), q) = 0$. But this in turn means that $\nu_e$ is a smooth manifold, i.e. a smooth hypersurface near $e$ because, for every $q$ in $\nu_e$ there is a neighbourhood $U$ now so that $U \to \mathbb{R}^4$ is smooth. Finally, recall that it has already been shown in course of the proof of part (b) of Lemma (Properties of $g$) that there cannot be a smooth hypersurface at $e$. $\square$

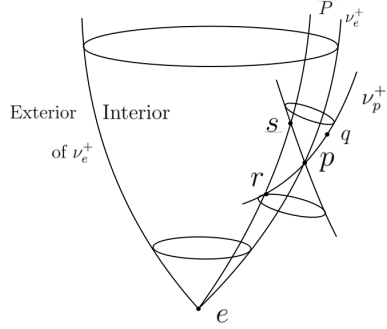

Figure 7 in Ehlers et al. (2012) (own drawing)

**Claim.** *(Null hypersurfaces) The integral curves associated to a null vector field $K$ that itself generates a null hypersurface $S$ are $\mathcal{C}$-null geodesics.*

*Proof.* We need to show that for such a vector field $K$, the geodesic equation (up to reparameterisation) $\nabla_K K = \lambda K$ obtains. Importantly, this statement follows from the fact that at each $p \in S$, we have $\nabla_K K \perp T_p S$ (call this condition $(*)$), that is $\mathcal{g}(\nabla_K K, X) = 0$ for all $X \in T_p S$. (After all, if $\nabla_K K$ is orthogonal to the hypersurface $S$, it can only be parallel to its defining orthogonal vector $K$.) We will now show $(*)$ and thus establish the claim.

First, extend the arbitrary tangent vector $X \in T_p S$ via a flow induced by $K$ such that $X$ is invariant under that flow, i.e. $[K, X] := \nabla_K X - \nabla_X K = 0$. As $X$ remains tangent to $T_p S$ along the flow generated by $K$, we have $\mathscr{G}(X, K) = 0$.[20] One now differentiates with respect to $K$:

$$0 = K(\mathscr{G}(K, X)) = (\nabla_K \mathscr{G})(K, X) + \mathscr{G}(\nabla_K K, X) + \mathscr{G}(K, \nabla_K X).$$

Now, a conformal connection can be induced from the Levi-Civita connection associated with an arbitrary element of a conformal equivalence class of metrics (this follows in particular in the current context of interest from the algebraic definition of conformal structure of $\mathcal{C}$) ; for such a conformal connection, one has $\nabla_\mu \mathscr{G}_{\nu\lambda} = 0$ (Curry and Gover, 2018, p. 17). Thus, from the above, one has $0 = \mathscr{G}(\nabla_K K, X) + \mathscr{G}(K, \nabla_K X)$, and hence

$$\mathscr{G}(\nabla_K K, X) = -\mathscr{G}(K, \nabla_K X) = -\mathscr{G}(K, \nabla_X K) = -\frac{1}{2} X(\mathscr{G}(K, K)) = 0,$$

where we used $[K, X] := \nabla_K X - \nabla_X K = 0$ in the second step. $\qquad\square$

**Proposition** (Characterisation of conformal structure $\mathcal{C}$). *Light rays are $\mathcal{C}$-null geodesics. $\nu_e$ is a $\mathcal{C}$-null cone at $e \in M$ then, and the map $e \mapsto \nu_e$ establishes a conformal structure on $M$.*

*Proof.* Choose a neighbourhood around a point $e$ such that (1) this neighbourhood is a $V$-type neighbourhood (see axiom $L_1$), and such that (2) it admits a timelike vector field, i.e., the neighbourhood is a time-orientable neighbourhood. (The existence of individual neighbourhoods of this kind is guaranteed from the previous claim and $L_1$; the chosen neighbourhood is just their joint intersection.)

Let $p$ be an event on the future light cone emanating from $e$ (itself denoted by $\nu_e^+$), within the $V$ neighbourhood. Without loss of generality, we can furthermore restrict $V$ to a neighbourhood as described in axiom $C$, i.e. ($*$) for each event $p$ in that neighbourhood, there is a particle through $e$ that also goes through $p$ if and only if $p$ lies in the interior of $\nu_e$.

The tangent space on the light cone $\nu_e^+$ at point $p \in V$ cannot be spacelike: if so, the cone could not be completely be filled with particles as, however, required by axiom $C$ (for recall from lemma (Properties of $\mathcal{P}$-geodesics, 1) that particles are nowhere spacelike).

Furthermore, the tangent space at $p$ on $\nu_e^+$ cannot be timelike: assume that the tangent space is timelike. Consequently, the light cone $\nu_p^+$ will intersect—and not align with—$\nu_p^+$ at $p$. Note then that one can find a light ray $(r, p, q)$ that goes from some point $r$ in the interior of $\nu_e^+$ via $p$ on $\nu_e^+$ to some point $q$ in the exterior of $\nu_e^+$ (see figure 7). By choosing $r$ close enough to $p$, and using that according to axiom $C$ the interior of $\nu_e^+$ is filled up locally with particles passing through $e$,[21] one can find a particle $P$ from $e$ through $r$ that is connectible to $p$ through at least three (rather than exactly two) light rays (in clash with the initial $V$ neighbourhood assumption): by the two light rays arising from the intersection of the backward cone $\nu_p^-$ with $\nu_e^+$, and (at least) a light ray going from $p$ to $s$ (the latter is guaranteed from taking

---

[20] We follow up to this point the proof of Galloway (2004), proposition 3.1.

[21] In fact, EPS reference here axiom $P_2$ but it is clear that axiom $C$ must have been meant.

a $V$ neighbourhood within the assumed $V$ neighbourhood, that, however, excludes $e$).

This means then that the tangent space $T_p \nu_e^+$ for any $p$ in the specified neighbourhood of $e$ is null. Thus, $\nu_e^+$ is a smooth *null* hypersurface near $e$. As the null vector field associated to the light rays generates the hypersurface,[22] they must then be, in virtue of the proven claim (Null hypersurfaces), null geodesics. In particular, the local light cone $\nu_e$ is then indeed identical with the $\mathcal{C}$-null cone at $e$. $\qquad\square$

## 5   Free fall and projective structure

Having constructed a conformal structure $\mathcal{C}$ from axioms regarding the behaviour of light rays, EPS turn next to the construction of projective structure. They begin with two axioms:

**Axiom** ($P_1$). *Given an event $e$ and a $\mathcal{C}$-timelike direction $D$ at $e$, there exists one and only one particle $P$ passing through $e$ with direction $D$.*

**Axiom** ($P_2$). *For each event $e \in M$, there exists a coordinate system $(\overline{x}^\mu)$, defined in a neighbourhood of $e$ and permitted by the differential structure introduced in axiom $D_3$, such that any particle $P$ through $e$ has a parameter representation $\overline{x}^\mu(\overline{u})$ with*

$$\frac{d^2 \overline{x}^\mu}{d\overline{u}^2}\bigg|_e = 0;$$

*such a coordinate system is said to be projective at $e$.*

Regarding axiom $P_2$, Sklar (1977) raises the concern that this presupposes an understanding of which particles are forced versus force-free—since the force-free particles will satisfy the equation stipulated in the axiom, while forced particles will not—and that this understanding thereby transcends the empiricist motivations of EPS. There are two central lines of response here. The first is to accept that this a conventional, rather than empirically-motivated, input in the EPS scheme. The second is to argue that we are not, after all, free to stipulate which particles are forced versus force-free: for example, Coleman and Korté (1980) argue that the distinction can be *derived* merely from differential-topological inputs of the kind already discussed above. The issues here are delicate—for example, Pitts (2016) responds that the Coleman-Korté line of reasoning does not succeed. Suffice it here simply to register that this is a point of controversy: these matters are discussed in greater detail in the sequel paper (Part II).

Note also that if the set of allowed coordinate charts does not exclude holonomic coordinates, then torsion-freeness does not follow. EPS appear to assume implicitly that coordinate systems *are* holonomic (this will become particularly evident in the following corollary), thereby excluding torsionful spacetimes by fiat. (For further discussion on anholonomic coordinates, see Iliev (2006) and Hartley (1995).)

---

[22] Note that integral curves to a null vector field need not generate a hypersurface that is null; this follows from the claim (Null hypersurfaces) according to which integral curves of a null vector field linked to a null hypersurface are null geodesics, and the simple fact that not all null curves that are integral curves of a null vector field are null geodesics! However, we have proven that the generated hypersurface is indeed a null hypersurface.

**Corollary.** *In an arbitrary coordinate system, one has*

$$\ddot{x}^\mu + \Pi^\mu_{\ \nu\lambda}\dot{x}^\nu\dot{x}^\lambda = \lambda\dot{x}^\mu, \tag{1}$$

*where $\lambda$ depends on the choice of $u$ and where one can restrict $\Pi^\mu_{\ \nu\lambda}$ to $\Pi^\mu_{\ [\nu\lambda]} = 0$ and $\Pi^\mu_{\ \nu\mu} = 0$.*

Note here that restriction to symmetric connection coefficients explicitly rules out torsionful spacetimes, as discussed above. Furthermore, note that imposing $\Pi^\mu_{\ \nu\mu} = 0$ is always justified for a projective connection, since for any projective connection for which this does not hold, one can always find an equivalent projective connection for which the condition does hold (i.e., the equivalent connection defines the same paths)—see (Crampin and Saunders, 2007, p. 697).

**Definition** (Projective structure (EPS)). *The structure imposed using $\Pi^\mu_{\ \nu\lambda}$ is known as a* projective structure $\mathcal{P}$: *a family of curves, called geodesics, whose members behave in the second-order infinitesimal of each neighbourhood of M like the straight lines of an ordinary projective four-space.*

**Definition** (Geodesic). *Any curve that satisfies (1) is said to be ($\mathcal{P}$)-geodesic.*

**Proposition** (Uniqueness of projective structure). *The coefficients $\Pi^\mu_{\ \nu\lambda}$ are determined uniquely at $e$ for any particular coordinate system $\{x^\mu\}$ around $e$.*

*Proof.* Suppose there is another projective structure $\overline{\Pi}^\mu_{\ \nu\lambda}$ determined at $e$ in this coordinate system by an analogous equation to (1), that is

$$\ddot{x}^\mu + \overline{\Pi}^\mu_{\ \nu\lambda}\dot{x}^\nu\dot{x}^\lambda = \lambda\dot{x}^\mu.$$

Denote the difference between $\overline{\Pi}^\mu_{\ \nu\lambda}$ and $\Pi^\mu_{\ \nu\lambda}$ as $\tilde{\Delta}^\mu_{\ \nu\lambda}$. Note that $\tilde{\Delta}^\mu_{\ \nu\lambda}$ is symmetric in its lower two indices, since $\overline{\Pi}^\mu_{\ \nu\lambda}$ and $\Pi^\mu_{\ \nu\lambda}$ are both also symmetric in their lower two indices.

Now, axiom $\mathcal{P}_1$ and equation (1) imply $T^{[\lambda}\tilde{\Delta}^{\mu]}_{\ \nu\rho}T^\nu T^\rho = 0$ where $T^\lambda := \dot{x}^\lambda$,i.e. is the tangent to the curve. To derive $\tilde{\Delta}^\lambda_{\ \mu\rho} = 0$, proceed as follows:

For all $T^\mu$, it holds that

$$T^{[\lambda}\tilde{\Delta}^{\mu]}_{\ vd}T^\nu T^\rho = 0$$
$$\Rightarrow \left(T^\lambda\tilde{\Delta}^\mu_{\ (\nu\rho)} - T^\mu\tilde{\Delta}^\lambda_{\ (\nu\rho)}\right)T^\nu T^\rho = 0$$
$$\Rightarrow T^\lambda\tilde{\Delta}^\mu_{\ (\nu\rho)} - T^\mu\tilde{\Delta}^\lambda_{\ (\nu\rho)} = \Xi^{\lambda\mu}_{(\nu\rho)} \quad \text{and} \quad \Xi^{\lambda\mu}_{(\nu\rho)}T^\nu T^\rho = 0$$
$$\Rightarrow T^\lambda\tilde{\Delta}^\mu_{\ (\nu\rho)} - T^\mu\tilde{\Delta}^\lambda_{\ (\nu\rho)} = 0$$
$$\Rightarrow \tilde{\Delta}^\mu_{\ (\nu\rho)} = 0$$

The second to last step follows from the fact that $\Xi^{\lambda\mu}_{(\nu\rho)}$ would have to be linear in $T^\mu$ or $T^\lambda$, thus $\Xi^{\lambda\mu}_{(\nu\rho)} = \alpha T^\lambda\Xi'^\mu_{(\nu\rho)} + \beta T^\mu\Xi'^\lambda_{(\nu\rho)}$ for some factors $\alpha, \beta$. Now, for any $T^\mu$, the second equation in the third line above becomes $\alpha\Xi'^\mu_{(\nu\rho)}T^\lambda T^\nu T^\rho + \beta\Xi'^\lambda_{(\nu\rho)}T^\mu T^\nu T^\rho = 0$, with $\Xi'^\mu_{(\nu\rho)}$ independent of $T^\mu$. From that $\Xi'^\mu_{(\nu\rho)}$ is independent of $T^\mu$, we can infer that $\Xi'^\mu_{(\nu\rho)}$ must be zero, in which case the penultimate line follows. The last step follows from that $\Delta^\mu_{\ (\nu\rho)}$ is independent of $T^\mu$.

Thus, the antisymmetric and symmetric parts both vanish, and we have $\tilde{\Delta}^\mu_{\ \nu\lambda} = 0$. $\qquad\square$

**Claim.** *The coefficients $\Pi^\mu_{\;\nu\lambda}$ can be measured in radar coordinates.*

*Proof.* First, manipulate (1) by contracting it with another tangent vector while antisymmetrising it:

$$\ddot{x}^\mu + \Pi^\mu_{\;\nu\lambda}\dot{x}^\nu\dot{x}^\lambda = \lambda\dot{x}^\mu$$

$$\Rightarrow \dot{x}^{[\mu}(\ddot{x}^{\sigma]} + \Pi^{\sigma]}_{\;\;\nu\lambda}\dot{x}^\nu\dot{x}^\lambda) = \lambda\dot{x}^{[\mu}\dot{x}^{\sigma]}$$

$$\Rightarrow \dot{x}^{[\mu}(\ddot{x}^{\sigma]} + \Pi^{\sigma]}_{\;\;\nu\lambda}\dot{x}^\nu\dot{x}^\lambda) = 0. \tag{2}$$

Note now that all $x^\mu$, $\dot{x}^\mu$, etc. can in principle be constructed using radar coordinates, while the parameter $\lambda$—which is not directly determinable empirically—has been removed, so the result follows. $\qquad\square$

Thus, EPS have: (i) derived a projective structure $\mathcal{P}$ from the salient empirically-motivated axioms, (ii) demonstrated the uniqueness of this structure, and (iii) demonstrated how this structure can be measured operationally using radar coordinates.

# 6 Compatibility of free fall and light propagation: the construction of Weyl structure

Having constructed a conformal structure $\mathcal{C}$ from the trajectories of light rays, and a projective structure $\mathcal{P}$ from the trajectories of freely-falling particles, EPS now introduce a 'compatibility' axiom in order to build a Weyl structure from these structures. This compatibility axiom is stated as follows:

**Axiom** (*C*). *Each event $e$ has a neighbourhood $U$ such that an event $p \in U$, $p \neq e$ lies on a particle $P$ through $e$ if and only if $p$ is contained in the interior of the light cone $\nu_e$ of $e$.*

In other words: For all events $e$, there exists a neighbourhood $U$ such that for all events $p \in U$ the following equivalence holds: there exists a particle $P$ with $p, e \in P$ iff $p$ lies in the interior of $\nu_e$.

From axiom $C$, EPS can prove the following Lemma:

**Lemma** (Properties of $\mathcal{P}$-geodesics).

1. *Every particle is a $\mathcal{P}$-geodesic which is nowhere $\mathcal{C}$-spacelike.*

2. *Every $\mathcal{P}$-geodesic that is timelike at some event can nowhere be spacelike.*

*Proof. ad (1):* That every particle is a $\mathcal{P}$-geodesic follows from axiom $P_2$. Assume for contradiction that the $\mathcal{P}$-geodesic describing a particle $P$ is spacelike at $e$. Tangent vectors, via the exponential map, uniquely define up to a certain parameter value $r$ curves that do not intersect with one another; in particular, the curve $\gamma$ associated via the (inverse) exponential map to the spacelike tangent at $e$ to $P$ (with $\{\gamma(t) : t \in [0, r)\} \subset P$) does not intersect with the curve $\eta$ associated to a null vector at $e$ as long as $r < r'$. Now, choose an $s$ such that $0 < s < r$. It then follows that the point $p_s := \gamma(s)$ will be on $P$ but not in the interior of $\nu_e$ (this would require having intersected with $\eta$).

But then there exists an event (namely $e$) such that for all its neighbourhoods $U$ there exists $p \in U$ (namely $p_s$ with sufficiently small $s > 0$) on a particle $P$ while $p \notin \text{int}(\nu_e)$—in clash with axiom $C$ that particles do not leave the cone.

*ad (2):* A $\mathcal{P}$-geodesic which is timelike at an event must be a particle. (For, there will also then be a 'particle' $\mathcal{P}$-geodesic at the same event with the same tangent; as a geodesic equation is a second-order equation, this means that the two $\mathcal{P}$-geodesics are identical.) From this, the result follows from part (1) of the lemma. $\qquad\square$

With this lemma established, EPS then provide a number of definitions:

**Definition.** *The inverse $\mathscr{g}^{\mu\nu}$ of the conformal metric density is given by*

$$\mathscr{g}^{\mu\nu}\,\mathscr{g}_{\nu\lambda} = \delta^{\mu}{}_{\lambda}.$$

**Definition.** *The components of the conformal connection are given by*

$$K^{\mu}{}_{\nu\lambda} := \frac{1}{2}\mathscr{g}^{\mu\sigma}\left(\mathscr{g}_{\sigma\nu,\lambda} + \mathscr{g}_{\sigma\lambda,\nu} - \mathscr{g}_{\nu\lambda,\sigma}\right).$$

**Definition.** *The difference between the projective and conformal connection is given by*

$$\Delta^{\mu}{}_{\nu\lambda} := \Pi^{\mu}{}_{\nu\lambda} - K^{\mu}_{\nu\lambda},$$
$$\Delta_{\mu\nu\lambda} := \mathscr{g}_{\mu\sigma}\Delta^{\sigma}{}_{\nu\lambda}.$$

**Definition.** *It is convenient to define the following auxiliary objects:*

$$p_{\mu} := -\frac{8}{9}\Delta_{[\mu\sigma]}{}^{\sigma},$$
$$L_{\mu\nu\lambda} := \frac{4}{3}\Delta_{[\mu\nu]\lambda} - p_{[\mu}\mathscr{g}_{\nu]\lambda}.$$

Having defined these quantities, EPS then prove two lemmas. The first regards the properties of $L_{\mu\nu\lambda}$:

**Lemma** (Properties of $L_{\mu\nu\lambda}$). *$L_{\mu\nu\lambda}$ satisfies*

$$L_{(\mu\nu)\lambda} = L_{[\mu\nu\lambda]} = L^{\mu}{}_{\nu\mu} = 0. \tag{3}$$

*Proof.* This follows by direct computation using the above-defined objects. $\qquad\square$

The second lemma regards the properties of $\Delta^{\mu}{}_{\nu\lambda}$:

**Lemma** ((a) Properties of $\Delta^{\mu}{}_{\nu\lambda}$). *$\Delta^{\mu}{}_{\nu\lambda}$ satisfies the following results:*

1. $\Delta^{\mu}{}_{[\nu\lambda]} = \Delta^{\mu}{}_{\nu\mu} = 0.$

2. $\Delta_{\mu\nu\lambda} = \Delta_{(\mu\nu\lambda)} + \frac{1}{2}(p_{\mu}\mathscr{g}_{\nu\lambda} - \mathscr{g}_{\mu(\nu}p_{\lambda)}) + L_{\mu(\nu\lambda)}.$

*Proof. ad (1):* For $\Delta^{\mu}{}_{[\nu\lambda]} = 0$, we have already seen that $\Pi^{\mu}{}_{[\nu\lambda]} = 0$, so it suffices to show that $K^{\mu}{}_{[\nu\lambda]} = 0$. To demonstrate this, consider

$$K^{\mu}{}_{[\nu\lambda]} = \frac{1}{2}\mathscr{G}^{\mu\sigma}(\mathscr{G}_{\sigma[\nu,\lambda]} + \mathscr{G}_{\sigma[\lambda,\nu]} - \mathscr{G}_{[\nu\lambda],\sigma}).$$

The final term on the RHS here vanishes by the symmetry of $\mathscr{G}_{\mu\nu}$; the first two terms on the RHS cancel; thus, the result follows.

For $\Delta^{\mu}{}_{\nu\mu} = 0$, consider

$$K^{\mu}{}_{\nu\mu} = \frac{1}{2}\mathscr{G}^{\mu\sigma}(\mathscr{G}_{\sigma\nu,\mu} + \mathscr{G}_{\sigma\mu,\nu} - \mathscr{G}_{\nu\mu,\sigma}) = 0;$$

since we have already argued above that for a projective connection one can take $\Pi^{\mu}{}_{\nu\mu} = 0$, the result follows.

*ad (2):* This result follows from the stated definitions of the relevant objects, and the decomposition of a rank three tensor into irreducible representations.

<div align="right">□</div>

With all of these results established, EPS then prove the following claim, the purpose of which is to learn more regarding the relationship between the projective structure $\mathcal{P}$ and the conformal structure $\mathcal{C}$:

**Claim** ($\mathcal{P}$-geodesic with tangent $\mathcal{C}$-vector at a point)**.** *Let $x^{\mu}(u)$ describe a $\mathcal{P}$-geodesic such that $\dot{x}^{\mu}(0)$ is a $\mathcal{C}$-null vector. Then, at $u = 0$,*

$$\frac{d}{du}(\mathscr{G}_{\mu\nu}\dot{x}^{\mu}\dot{x}^{\nu}) = -\Delta_{\mu\nu\lambda}\dot{x}^{\mu}\dot{x}^{\nu}\dot{x}^{\lambda}.$$

*Proof.* One can establish this result by manipulating the geodesic equation as follows:

$$\ddot{x}^{\mu} + \Pi^{\mu}{}_{\nu\lambda}\dot{x}^{\nu}\dot{x}^{\lambda} = \lambda\dot{x}^{\mu}$$

$$\Rightarrow \ddot{x}^{\mu} + \Delta^{\mu}{}_{\nu\lambda}\dot{x}^{\nu}\dot{x}^{\lambda} + K^{\mu}{}_{\nu\lambda}\dot{x}^{\nu}\dot{x}^{\lambda} = \lambda\dot{x}^{\mu}$$

$$\Rightarrow \ddot{x}^{\mu} + \Delta^{\mu}{}_{\nu\lambda}\dot{x}^{\nu}\dot{x}^{\lambda} + \frac{1}{2}\mathscr{G}^{\mu\sigma}\left(\partial_{\lambda}\mathscr{G}_{\sigma\nu} + \partial_{\nu}\mathscr{G}_{\sigma\lambda} - \partial_{\sigma}\mathscr{G}_{\nu\lambda}\right)\dot{x}^{\nu}\dot{x}^{\lambda} = \lambda\dot{x}^{\mu}$$

$$\Rightarrow \mathscr{G}_{\delta\mu}\ddot{x}^{\mu} + \mathscr{G}_{\delta\mu}\Delta^{\mu}{}_{\nu\lambda}\dot{x}^{\nu}\dot{x}^{\lambda} + \frac{1}{2}\mathscr{G}_{\delta\mu}\mathscr{G}^{\mu\sigma}\left(\partial_{\lambda}\mathscr{G}_{\sigma\nu} + \partial_{\nu}\mathscr{G}_{\sigma\lambda} - \partial_{\sigma}\mathscr{G}_{\nu\lambda}\right)\dot{x}^{\nu}\dot{x}^{\lambda} = \lambda\,\mathscr{G}_{\delta\mu}\dot{x}^{\mu}$$

$$\Rightarrow \mathscr{G}_{\delta\mu}\ddot{x}^{\mu} + \Delta_{\delta\nu\lambda}\dot{x}^{\nu}\dot{x}^{\lambda} + \frac{1}{2}\left(\partial_{\lambda}\mathscr{G}_{\delta\nu} + \partial_{\nu}\mathscr{G}_{\delta\lambda} - \partial_{\delta}\mathscr{G}_{\nu\sigma}\right)\dot{x}^{\nu}\dot{x}^{\lambda} = \lambda\,\mathscr{G}_{\delta\mu}\dot{x}^{\mu}$$

$$\Rightarrow \mathscr{G}_{\delta\mu}\ddot{x}^{\mu}\dot{x}^{\delta} + \Delta_{\delta\nu\lambda}\dot{x}^{\nu}\dot{x}^{\lambda}\dot{x}^{\delta} + \frac{1}{2}\partial_{\lambda}\mathscr{G}_{\delta\nu}\dot{x}^{\nu}\dot{x}^{\lambda}\dot{x}^{\delta} = \lambda\,\mathscr{G}_{\delta\mu}\dot{x}^{\mu}\dot{x}^{\delta}$$

$$\Rightarrow \frac{d}{du}\left(\mathscr{G}_{\delta\mu}\dot{x}^{\mu}\dot{x}^{\delta}\right) + \Delta_{\delta\nu\lambda}\dot{x}^{\nu}\dot{x}^{\lambda}\dot{x}^{\delta} = 0$$

where we have used in the penultimate line that $\dot{x}^{\mu}$ is null at $u = 0$ (allowing us for dropping the RHS), and have rescaled $\Delta_{\delta\nu\lambda}$ by a constant factor.

<div align="right">□</div>

Using this claim, one can then prove the following lemma:

**Lemma** ((b) Property of $\Delta^{\mu}{}_{\nu\lambda}$)**.** $\Delta_{(\mu\nu\lambda)} = g_{(\mu\nu}s_{\lambda)}$ *identically.*

*Proof.* If the expression $\Delta_{\mu\nu\lambda}\dot{x}^\mu\dot{x}^\nu\dot{x}^\lambda = 0$ is non-zero for a vector $\dot{x}^\mu$ at $u = 0$, then there is a change from a timelike (spacelike) vector for $u < 0$ to a spacelike (timelike) vector for $u > 0$. This contradicts the first lemma derived from axiom $C$ above, according to which every particle which is a $\mathcal{P}$-geodesic is nowhere spacelike.

But if $\Delta_{\mu\nu\lambda}\dot{x}^\mu\dot{x}^\nu\dot{x}^\lambda = 0$ holds for a null vector, then $\Delta_{\mu\nu\lambda}\dot{x}^\mu\dot{x}^\nu\dot{x}^\lambda$ must be a function $f$ of $g_{\mu\nu}\dot{x}^\mu\dot{x}^\nu$. One way to realise this concretely is to set $\Delta_{(\mu\nu\lambda)} = g_{(\mu\nu}s_{\lambda)}$ for arbitrary one-form $s_\lambda$ such that $\Delta_{\mu\nu\lambda}\dot{x}^\mu\dot{x}^\nu\dot{x}^\lambda = (g_{\mu\nu}\dot{x}^\mu\dot{x}^\nu)s_\lambda\dot{x}^\lambda$. Notably, the choice of $\Delta^\mu{}_{\nu\lambda}$ is unique since $\Delta^\mu{}_{\nu\lambda} = \Pi^\mu{}_{\nu\lambda} - K^\mu{}_{\nu\lambda}$ and neither the connections $\Pi^\mu{}_{\nu\lambda}$ nor $K^\mu{}_{\nu\lambda}$ are functions of tangent vectors. □

EPS are then in position to establish the first results on the relation between $\mathcal{P}$ and $\mathcal{C}$:

**Claim** (Relation between $\mathcal{P}$ and $\mathcal{C}$). *On any $\mathcal{P}$-geodesic $x^\mu(u)$, the following relation obtains:*

$$\frac{d}{du}(\mathscr{g}_{\mu\nu}\dot{x}^\mu\dot{x}^\nu) = (\mathscr{g}_{\mu\nu}\dot{x}^\mu\dot{x}^\nu)(2\lambda - s_\lambda\dot{x}^\lambda).$$

*Proof.* Consider the penultimate line of the proof of the above claim. Inserting the identity from the above lemma, one obtains the desired result. □

**Corollary** (Relation between $\mathcal{P}$ and $\mathcal{C}$). *A $\mathcal{P}$-geodesic that is timelike, spacelike, or null respectively, with respect to $\mathcal{C}$ at one of its events, has the same orientation everywhere.*

*Proof.* Let $f(u) := g_{\mu\nu}\dot{x}^\mu\dot{x}^\nu$. If $\dot{x}^\mu$ is a null vector for some $u$, then since $\dot{f} = 0$ in light of the previous claim (Relation between $\mathcal{P}$ and $\mathcal{C}$), $\dot{x}^\mu$ is null for all $u$ (by the above). If $\dot{x}^\mu$ is timelike or spacelike at some parameter time $u_0$, we find upon re-ordering into $\dot{f}/f = 2\lambda - s_\lambda\dot{x}^\lambda$ and logarithmic integration that

$$\log f(u) - \log f(u_0) = \int_{u_0}^u du'(2\lambda - s_\lambda\dot{x}^\lambda),$$

and hence that

$$f(u) = f(u_0)\cdot\exp(...),$$

i.e. as $\exp(...) > 0$ the sign of $f(u)$ for any $u$ is the same as for $f(u_0)$. □

One can now define:

**Definition** (EPS-compatibility between projective and conformal structure). *A projective structure and a conformal structure are* EPS-compatible *iff the null geodesics of the conformal structure are also projective geodesics (but not necessarily vice versa). (See Trautman (2012).)*

Before showing EPS-compatibility between $\mathcal{P}$ and $\mathcal{C}$, the following lemma is established:

**Lemma.** *Each event $p$ on $\nu_e$ sufficiently close to $e$ can be approximated arbitrarily closely by events $q$ situated on particles through $e$.*

*Proof.* Axiom $C$ states that for any event $e$ there is a neighbourhood such that all events within that neighbourhood *that are connected to $e$ via particles* will lie inside $\nu_e$. If $p$ is sufficiently close to $e$, then it is sufficiently close to particles $q$ inside $\nu_e$ that are connected to $e$ via particles. □

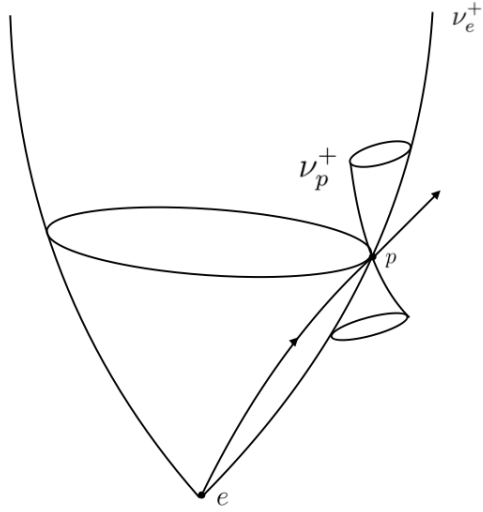

Figure 9 in Ehlers et al. (2012) (own drawing)

**Proposition** (EPS-compatibility). $\mathcal{C}$ *and* $\mathcal{P}$ *are EPS-compatible.*

*Proof.* Let $p \in \nu_e$, $p \neq e$. Let $q_n$ be a sequence of events within a $\mathcal{P}$-convex neighbourhood of $e$ that together with $e$ respectively uniquely define particle geodesics $P_n$,[23] each particle $P_n$ can be taken to obey a (projective) geodesic equation $\ddot{x} + \Pi^\lambda_{\mu\nu}\dot{x}^\mu\dot{x}^\nu = \lambda_n\dot{x}$ parameterised in such a way that for $\lambda_n = 0$ the geodesic goes through $e$, and for $\lambda_n = 1$ the geodesic goes through $q_n$.

The sequence $\{T_n\}$ of the tangent vectors of the $P_n$ at $e$ converges to the tangent vector $T$ of that geodesic $P$ which passes for $u = 0$ and $u = 1$ respectively through $e$ and $p$—provided $p$ lies in a sufficiently small $\mathcal{P}$-convex coordinate neighbourhood of $e$.

For $T_n \to T$ with $T_n$ non-spacelike, $T$ is either timelike or null: for any tangent vector $S$, the map $S^\mu \mapsto g_{\mu\nu}S^\mu S^\nu$ is a continuous function, i.e. convergence of a sequence $S^\mu$ must coincide with convergence of $g_{\mu\nu}S^\mu S^\nu$. As for all $T_n$, $g_{\mu\nu}T_n^\mu T_n^\nu < 0$, thereby $g_{\mu\nu}T^\mu T^\nu \leq 0$. At the same time, we can exclude that $T$ is timelike: for if it were timelike, then $P$—coming from $e$—would have had to intersect $\nu_e^+$ and would therefore at some point have to be spacelike, which is disallowed.

We now want to show that, between passing $e$ and $p$, $P$ can only contain events on $\nu_e$. First, $q \in P$ cannot be in the exterior of $\nu_e$ between passing $e$ and $p$. Secondly, assume $q \in P$ was in the interior of $\nu_e$ between passing $e$ and $p$. Then: $q \in \nu_q^+$; but at least locally, $\nu_q^+ \subset \nu_e^+$. Then a sufficiently close $p$

---

[23]A neighbourhood $V$ is $\mathcal{P}$-convex iff, for any two points $p, q \in V$, there exists exactly one geodesic segment $\gamma$ fully within $V$ that also contains $p$ and $q$. Cf. (Perlick, 2008, p. 134).

could not be reached by $P$ anymore—even though $p \in P$. Thus, $q$ must be on $\nu_e$.

To summarise: (1) the tangents along $P$ are null between $e$ and $p$ (what we said for the tangent vector at $p$ holds without any restriction for the tangents at the points to $P$ between $e$ and $p$ too). $P$ is thereby a projective null geodesic. (2) Between $e$ and $p$, $P$ is contained in $\nu_e$; for this region then, $P$ is a $\mathcal{C}$-null geodesic. Given that being a geodesic is a local property and that the considerations above hold around any other event $e$, it follows that for every null geodesic a projective null geodesic can be constructed that coincides with it. $\qquad\square$

It is worth pointing out that EPS claim in their main text (at the bottom of p. 80) to show that "projective null geodesics and conformal null geodesics are identical". Note that they only show that every $\mathcal{C}$-null geodesic is a projective (null) geodesic (EPS-compatibility)—and not its converse—which is, however, sufficient for all what follows below.

EPS-compatibility between $\mathcal{C}$ and $\mathcal{P}$ has two important corollaries:

**Corollary.** *For $\mathcal{P}$ and $\mathcal{C}$, the following condition holds:*

$$\Delta^{\mu}_{\ \nu\lambda} = 5q^{\mu}\mathscr{G}_{\nu\lambda} - 2\delta^{\mu}_{\ (\nu}q_{\lambda)} \tag{4}$$

*where $q_{\lambda}$ is a one-form.*

*Proof.* $\mathscr{G}_{\mu\nu}\dot{x}^{\mu}\dot{x}^{\nu} = 0$ and $\ddot{x}^{\mu} + K^{\mu}_{\ \nu\lambda}\dot{x}^{\mu}\dot{x}^{\nu} = \nu\dot{x}^{\mu}$ together characterise $\mathcal{C}$-null geodesics; as these are, however, also projective geodesics, they may also be described as $\ddot{x}^{\mu} + \Pi^{\mu}_{\ \nu\lambda}\dot{x}^{\nu}\dot{x}^{\lambda} = \lambda\dot{x}^{\mu}$. Subtracting the second from the third equation gives $\Delta^{\mu}_{\ \nu\lambda}\dot{x}^{\nu}\dot{x}^{\lambda} = (\lambda - \nu)\dot{x}^{\mu}$. Using the second part of Lemma ((a) Properties of $\Delta^{\mu}_{\ \nu\lambda}$), i.e., $\Delta_{\mu\nu\lambda} = \Delta_{(\mu\nu\lambda)} + \frac{1}{2}(p_{\mu}\mathscr{G}_{\nu\lambda} - \mathscr{G}_{\mu(\nu}p_{\lambda)}) + L_{\mu(\nu\lambda)}$, together with $\Delta^{\mu}_{\ [\nu\lambda]} = \Delta^{\mu}_{\ \nu\mu} = 0$ and $L_{(\mu\nu)\lambda} = L_{[\mu\nu\lambda]} = L^{\mu}_{\ \nu\mu} = L^{\mu}_{\ \nu\mu} = 0$ gives $\Delta^{\mu}_{\ \nu\lambda} = L^{\mu}_{\ (\nu\lambda)} + 5q^{\mu}\mathscr{G}_{\nu\lambda} - 2\delta^{\mu}_{\ (\nu}q_{\lambda)}$. From this (and the previous identity for $\Delta^{\mu}_{\ \nu\lambda}$), however, we obtain, for null curves with tangent $T^{\mu}$,

$$L^{\mu}_{\ (\nu\lambda)}T^{\nu}T^{\lambda} - 2\delta^{\mu}_{\ (\nu}q_{\lambda)}T^{\nu}T^{\lambda} = (\lambda - \nu)T^{\mu}.$$

As $2\delta^{\mu}_{\ (\nu}q_{\lambda)}T^{\nu}T^{\lambda} = 2q_{\lambda}T^{\lambda}T^{\mu}$, we have

$$L^{\mu}_{\ \nu\lambda}T^{\nu}T^{\lambda} = L^{\mu}_{\ (\nu\lambda)}T^{\nu}T^{\lambda} \propto T^{\mu}.$$

As, again, the latter equation holds for any null vector $T^{\mu}$, it follows that $L^{\mu}_{\ (\nu\lambda)} = 0$, so the stated result follows.

$\qquad\square$

**Corollary.** *The set of all particles is identical to the set of all $\mathcal{C}$-timelike $\mathcal{P}$-geodesics.*

*Proof.* We already know that particles are $\mathcal{P}$-geodesics, so what is at stake is whether they are always $\mathcal{C}$-timelike. If a particle were at any point $p$ a $\mathcal{C}$-null geodesic, then it would, in light of the claim previously proven, have to be a $\mathcal{P}$-null geodesic at $p$ (as implied by EPS compatibility), and, in light of corollary (relation between $\mathcal{P}$ and $\mathcal{C}$), then everywhere. But particles are not just $\mathcal{C}$-null given that a particle is, by definition, a geodesic that is timelike at *some* event (see bottom of p. 77 in EPS). This then also rules out that a particle could be $\mathcal{C}$-spacelike: it would have to be $\mathcal{C}$-timelike at some point and thus, by continuity, $\mathcal{C}$-null at some point. Thus, $\mathcal{P}$-geodesics must be $\mathcal{C}$-timelike geodesics everywhere. □

These results on EPS-compatibility give rise to what we call an *EPS-Weyl space*:

**Definition** (EPS-Weyl space). *A manifold M together with a EPS-compatible pair of conformal structure $\mathcal{C}$ and projective structure $\mathcal{P}$ is called a* EPS-Weyl *space* $(M, \mathcal{C}, \mathcal{P})$.

We can also define, following Scholz (2020), the notions of *Weyl structure* and *Weyl manifold*—and, relatedly, the notions of *Weyl connection* and *Weyl metric*:[24]

**Definition** (Weyl structure and Weyl manifold (Scholz 2020, definition 1)).

1. *A (pseudo-Riemannian)* Weyl structure *is given by the triple $(M, \mathcal{C}, \nabla)$ where M is a differentiable manifold, $\mathcal{C} = [g_{\mu\nu}]$ is a conformal class of pseudo-Riemannian metrics $g_{\mu\nu}$ on M, and $\nabla$ is the covariant derivative of a torsion free affine connection $\Gamma$ (called* Weyl *connection), constrained by the compatibility condition that for any $g_{\mu\nu} \in \mathcal{C}$ there is a differential 1-form $\varphi_\mu$ such that $\nabla_\lambda g_{\mu\nu} + 2\varphi_\lambda g_{\mu\nu} = 0$.*

2. *A* Weyl manifold $(M, [(g_{\mu\nu}, \varphi_\mu)])$ *is a differentiable manifold M endowed with a Weyl metric defined by an equivalence class of pairs $(g_{\mu\nu}, \varphi_\mu)$, where $g_{\mu\nu}$ is a pseudo-Riemannian metric on M and $\varphi_\mu$ is a (real valued) differential 1-form on M. Equivalence is defined by conformal rescaling $g_{\mu\nu} \mapsto \tilde{g}_{\mu\nu} = \Omega^2 g_{\mu\nu}$ and the corresponding transformation $\varphi_\mu \mapsto \varphi_\mu - d_\mu \ln \Omega$.*

EPS' argument for the claim that an EPS-Weyl space in fact gives rise to a Weyl structure/Weyl manifold is highly schematic and non-rigorous; we thus base our following presentation of how EPS establish a Weyl structure upon the excellent article by Matveev and Scholz (2020).[25]

Notably, both structures—i.e., that of a Weyl structure and that of a Weyl manifold—are equivalent (see remarks below definition 1 of Scholz 2020); we will from now on use both notions interchangeably. Note also then that the family of $\varphi_\mu$ defining a Weyl structure/Weyl manifold differ only by exact forms; the common exterior derivative is called distant curvature ('Streckenkrümmung'), denoted by $F_{\mu\nu}$. That is, $F_{\mu\nu}$ is given by $F_{\mu\nu} = d_\mu \varphi_\nu = d_\mu \tilde{\varphi}_\nu = \ldots$[26]

**Claim** (Existence of a Weyl connection). *The expression*

$$\Gamma^\mu_{\nu\lambda} = K^\mu_{\nu\lambda} + 5q^\mu \mathscr{g}_{\nu\lambda} - 10\delta^\mu_{(\nu} q_{\lambda)}$$

*defines a Weyl connection on the EPS-Weyl space $(M, \mathcal{C}, \mathcal{P})$ where $q_\lambda$ is a one-form determined in terms of $\mathcal{C}$ and $\mathcal{P}$ via equation (4).*

---

[24] Note that the following definitions follow Scholz (2020) closely, with only minor modifications to terminology.

[25] For a concise pedagogical introduction to Weyl geometry—albeit with a slightly different (but nevertheless illuminating) terminology to that used Matveev and Scholz (2020)—see Folland (1970).

[26] Strictly speaking, the distant curvature is only well-motivated along these lines if exact forms are closed. A sufficient condition which will be required from now for neighbourhoods is simply-connectedness.

*Proof.* See theorem 1 of Matveev and Scholz (2020). □

Given the equivalence of Weyl structure and Weyl manifold, this also defines a Weyl metric.

**Theorem** (Uniqueness of the Weyl metric ('Weyl's theorem'), cited after Scholz (2020))**.** *A Weyl metric is uniquely determined by its projective and conformal structures.*

*Proof.* See Weyl (1921). □

We have thus obtained an unique Weyl structure from $\mathcal{P}$ and $\mathcal{C}$, characterisable through the triple $(M, \mathcal{C}, \mathcal{A})$ where $\mathcal{A}$ is the affine structure associated to the Weyl connection. There has been an interesting recent discussion—promulgated in the critical introductory comments of Trautman (2012) to the 2012 reprint of EPS in *General Relativity and Gravitation* as a 'golden oldie'—as to whether the EPS condition of compatibility between conformal and projective structure is sufficient (and not just necessary) to establish the existence of a Weyl connection (see in particular Matveev and Trautman (2014)). Ultimately, as shown by Matveev and Scholz (2020), the compatibility condition stated by EPS is indeed sufficient—we refer readers to that paper for the details. Second, an insightful account of the history of Weyl's theorem and its subsequent appearance in the EPS framework is given by Scholz (2020).

# 7 From Weyl manifold to Lorentzian manifold

EPS can give the following characterisation of a Lorentzian manifold in terms of a Weyl manifold. A Weyl manifold is Lorentzian if and only if one of the following criteria is fulfilled:

**Equidistance criterion** "the proper times $t, t'$ on two arbitrary, infinitesimally close, freely falling particles $P, P'$ are linearly related (to first order in the distance) by Einstein-simultaneity; i.e., whenever $p_1, p_2, \dots$ is an equidistant sequence of events on $P$ (ticking of a clock) and $q_1, q_2, \dots$ is the sequence of events on $P'$ that are Einstein-simultaneous with $p_1, p_2, \dots$ respectively, then $q_1, q_2, \dots$ is (approximately) an equidistant sequence on $P'$ ..." (p. 69)

**Congruence criterion (no second-clock effect)** "consider parallel transport of a vector $V_p$ from a point $p$ to a point $q$ along two different curves $P, P'$. The resulting vectors, at $q$, $V_q$ and $V'_{q'}$, will be different, in general ... If and only if $V_q$ and $V'_q$ are congruent [have the same length] for all such figures, the Weyl geometry considered is in fact, [pseudo-]Riemannian." (p. 69)

In order to prove that a Weyl manifold is Lorentzian if and only if one of these criteria is fulfilled, one makes recourse to two lemmas:

**Lemma** (Curvature decomposition in Weyl structure). *The curvature tensor $R^\lambda_{\ \mu\varepsilon\rho}$ associated with a Weyl connection decomposes uniquely according to*

$$R^\lambda_{\ \mu\nu\rho} = \hat{R}^\lambda_{\ \mu\nu\rho} + \frac{1}{2}\delta^\lambda_{\ \mu}F_{\nu\rho},$$

*where*

$$g_{\sigma(\lambda}\hat{R}^\sigma_{\ \mu)\nu\rho} = 0 \quad , \quad F_{(\lambda\mu)} = 0,$$

*with F being the distant curvature as introduced above.*

*Proof.* A proof is given in Folland (1970), statement E, p. 151. □

**Lemma.** *(Congruence criterion for Lorentzian geometry) In any simply connected domain of M, there is a Lorentzian metric $g_{\mu\nu}$ that is compatible with the conformal structure $\mathcal{C}$ such that the Weyl connection is metric with respect to $g_{\mu\nu}$ iff the distant curvature $F_{\mu\nu} = 0$.*

*Proof. Left-to-right*: If $\Gamma$ is metric, then by definition $\nabla_\mu g_{\nu\lambda} = 0$, in which case $d\varphi = 0$, and so $F = 0$.

*Right-to-left*: If $F = 0$, then $d\sigma = 0$, in which case locally $\sigma_\mu = d_\mu\varphi$ for some scalar field $\varphi$. But in that case, one is working in an integrable Weyl geometry; and in such a case, one can perform a Weyl gauge transformation (i.e., transformations on $(g_{\mu\nu}, \varphi)$ which fixes the same Weyl geometry)

$$g_{\mu\nu} \mapsto \tilde{g}_{\mu\nu} := e^f g_{\mu\nu},$$
$$\varphi \mapsto \tilde{\varphi} := \varphi - f$$

for suitable function $f$ in order to transform to the *Riemann gauge* of the integrable Weyl geometry, in which $\tilde{\varphi} = 0$. One then finds that the metricity condition is satisfied (Scholz, 2020, p. 31). □

More rigorously then, the congruence criterion can be established as follows:

**Proposition.** *(Congruence criterion for Lorentzian geometry) In any simply connected domain of M, there is a Lorentzian metric $g_{\mu\nu}$ that is compatible with the conformal structure $\mathcal{C}$ such that the Weyl connection $\Gamma$ is metric with respect to $g_{\mu\nu}$ iff there is no second clock effect.*

*Proof.* The second clock effect in a Weyl geometry occurs iff there non-vanishing length curvature $F_{\mu\nu}$. This can be seen by explicitly comparing the result of parallel transporting the same starting vector via two different paths from a point $p$ to a point $q$: there will only be no change in length iff the length connection part of the Weyl metric $\sigma_\mu$ (as defined in the previous proof) is closed, i.e. $d_\mu\sigma_\nu = 0$. But this is again directly equivalent to $F_{\mu\nu} = 0$ (see Bell and Korté (2016)). □

The equidistance criterion comes about by unpacking further what $F_{\mu\nu} = 0$ amounts to:

**Proposition.** *(Relation between equidistance criterion and Lorentzian geometry)*
*In any simply connected domain of M, there is a Lorentzian metric $g_{\mu\nu}$ that is compatible with the conformal structure $\mathcal{C}$ such that the Weyl connection $\Gamma$ is metric with respect to $g_{\mu\nu}$ iff equidistant events on a particle P (i.e. geodesics) correspond to equidistant events on a particle P' as described in the equidistance criterion.*

*Proof. Right-to-left:* Let $U$ be the tangent vector of the first of two adjacent, affinely parameterised geodesics $P, P'$ of a geodesic congruence, and $V$ their connection vector. Consider the corresponding geodesic deviation equation

$$\hat{V}^\mu = \hat{R}^\mu{}_{\nu\rho\sigma} U^\nu U^\rho V^\sigma + \frac{1}{2} U^\mu F_{\nu\rho} U^\nu V^\rho. \tag{5}$$

with $g_{\sigma(\lambda}\hat{R}^\sigma{}_{\mu)\nu\rho} = 0$. (For a derivation of the geodesic equation for a torsion-free connection, see Reall (2021), equation (4.22) and its subsequent proof. The specific equation here follows then from splitting up the Riemann tensor as allowed by the above lemma (Curvature decomposition in Weyl structure).) The LHS is $\hat{\nabla}_U \hat{\nabla}_U V^\mu$, which is orthogonal to $U^\mu$ since

$$\begin{aligned}
g_{\mu\nu} U^\nu \left( \hat{\nabla}_U \hat{\nabla}_U V^\mu \right) &= g_{\mu\nu} \hat{\nabla}_U \left( U^\nu \hat{\nabla}_U V^\mu \right) \\
&= g_{\mu\nu} \hat{\nabla}_U \hat{\nabla}_U \left( U^\nu V^\mu \right) \\
&= \hat{\nabla}_U \hat{\nabla}_U \left( g_{\mu\nu} U^\nu V^\mu \right) \\
&= 0.
\end{aligned}$$

The first term on the RHS is also orthogonal to $U^\mu$: After multiplying with $g_{\mu\zeta} U^\zeta$, one obtains $g_{\mu\zeta} U^\zeta \hat{R}^\mu{}_{\nu\rho\sigma} U^\nu U^\rho V^\sigma = g_{\mu[\zeta} \hat{R}^\mu{}_{\nu]\rho\sigma} U^\zeta U^\nu U^\rho V^\sigma = 0$ where the first step uses $g_{\sigma(\lambda}\hat{R}^\sigma{}_{\mu)\nu\rho} = 0$, and the last step follows from contraction of two antisymmetric with two symmetric indices. At the same time, the second part on the RHS is parallel and not orthogonal to $U$ because it is a scalar multiplied by $U$ (and $U$ is non-null).

Now, if equidistant events on $P$ correspond to equidistant events on $P'$ as described in the equidistance criterion, "the connection vector $V$ ... can be chosen orthogonal to $P$ [i.e. to $U$.]" (p. 82), because shared equidistance between two curves implies orthogonality. So, choose $V$ to be orthogonal to $U$. We now hit the whole equation (5) with $g_{\mu\nu} U^\nu$: the LHS as well as the first term on the RHS of (5) must vanish, as they are orthogonal to $U$. If this is true for arbitrary $U$ and $V$, then $F$ has to vanish. From this via the previous lemma it follows that there is a Lorentzian metric $g_{\mu\nu}$ that is compatible with the conformal structure $\mathcal{C}$ such that $\Gamma$ is metric with respect to $g_{\mu\nu}$.

*Left-to-right:*[27] Suppose that $\Gamma$ is metric with respect to $g_{\mu\nu}$. Given this metric compatibility, locally one can find a Fermi normal coordinate system in which connection coefficients vanish along a given geodesic (in that coordinate neighbourhood).[28] Furthermore, in these coordinates, geodesics neighbouring the Fermi coordinate-defining geodesic approximately take the form of straight line trajectories due to continuity, i.e. satisfy $d^2 x^\mu / d\lambda^2 =$

---

[27] Note that EPS do not prove the left-to-right direction of the above proof on page 82 of their article.

[28] For background on Fermi normal coordinates, see Manasse and Misner (1963).

0 for affine parameter $\lambda$ up to an error (which is a function of how closely the neighbouring geodesic is chosen). Then, the coordinate-defining geodesic (which could have been chosen arbitrarily) and a sufficiently closely chosen geodesic approximately correspond to two straight lines in that coordinate system, from which it follows that Einstein synchrony approximately defines level surfaces of proper time connecting the two geodesics (as familiar from the flat spacetime case), and so the equidistance criterion approximately obtains. Notably, the error by which the equidistance criterion is violated is controlled, i.e. it can be made ever smaller the closer the second geodesic is chosen to the Fermi coordinate-defining one.  □

Notably, for several statements, the (not further justified) assumption of (local) "simply connectedness" was required. One such instance, as already pointed out, was in the definition of the notion of distant curvature.The assumption of simply connectedness as such is harmless in so far as it does not exclude any GR spacetimes: it has been shown that any non-simply connected region of an asymptotically flat spacetime must lie behind a horizon (see, for instance, Schleich and Witt (2010)), which then also means that any GR spacetime must be locally simply connected with respect to the manifold topology.[29]

Moreover, no matter which of the above two criteria is used, both can—it seems even must—be operationalised through the use of clocks (after all, in the equidistant criterion, one makes recourse to Einstein conventionality, and, in the congruence criterion, recourse to clocks). Clock constructions available at the time of writing of Ehlers et al. (2012) were only for Lorentzian but not for more general spacetimes, in particular not for Weyl structure (see Ehlers (1973b) for a review; among others, Marzke and Wheeler (1964), Desloge (1989) and Castagnino (1968) are relevant here). First and foremost, the work of Perlick (1987) (among others[30]) established how clocks can indeed be constructed as a natural addition to the EPS-scheme within the Weyl structure, making the above two criteria workable after all.

# 8  Evaluating the EPS scheme

The EPS scheme sets up the core kinematical structure $(M, g_{ab})$ of classical Lorentzian spacetime theories—including general relativity—from local statements regarding the propagation of matter and light rays. This construction process of the kinematical structure is additive ('constructivist', in the sense of Carnap (1967), as discussed above), so that the resulting kinematical structure $(M, g_{ab})$ must of necessity be consistent with the input axioms—it is built up linearly therefrom. Insofar as the axioms are justified empirically, the resulting theory will have been placed on firm empirical footing (this, indeed, was also Reichenbach's motivation for the pursuit of the constructive axiomatisation of physical theories—see Reichenbach (1969)). However, insofar as these axioms are not all justified empirically, the empirical status of the resulting structure is not fully guaranteed. As we have seen in our identification of several conventional inputs in the EPS approach, as well as apparently unjustified assumptions and restrictions, this is a more accurate representation of the status of EPS' result.

---

[29]Nevertheless, on Hawking, King and McCarthy's path topology, generic spacetimes are not even locally simply connected (see Low (2010)).

[30]See Avalos et al. (2018) for a rigorous account of how Perlick's clock construction settles the issue. See also Köhler (1978) and Chandra (1984) for other Weylian clock constructions.

There are many more points to be made regarding the foundational ramifications of the EPS axiomatisation. We close this paper with a set of comments on (1) limitations and insufficiencies, (2) associated amendments, and (3) overall readings of the EPS scheme. We evaluate construct*ive* approaches to spacetime theories more generally in the companion paper (Part II).

## 8.1 Limitations and insufficiencies

Over the course of this walkthrough, the following limitations and deficiencies of the EPS scheme have become evident:

**No dynamics:** As it is the purpose of the EPS axiomatisation to recover the *kinematical* structure of general relativity, the approach, of course, does not recover the field equations. Presumably, the thought would be that the dynamics can in principle be straightforwardly systematised once the kinematics is in place.[31]

**Locality restrictions:** The EPS scheme holds only locally; nothing about global topology of spacetime can be learned via this approach.

**Idealisations:** The differentiability assumptions in EPS are an important control handle over possible alternative kinematical settings; they are thus not at all the innocent idealisation assumption which they appear to be at first sight. Furthermore, the operational assumptions deployed by EPS are highly idealised but—impractical idealisations granted—workable.

**Non-constructivist:** The EPS scheme arguably deviates from its basic posit of semantic linearity at different points and to different degrees of severity, i.e., strictly speaking the EPS system is thus not a constructivist system in the strict sense. These will be outlined in detail below.

Let us consider in more detail the sense in which the EPS axiomatisation is non-constructivist. First, the original scheme confirms a common stereotype on how constructivist rational reconstructions are *produced*: if stuck, break your own rule of linearity and simply postulate an *ad hoc* workaround. The bridging criteria from a Weyl metric to a Lorentzian metric offered by EPS—no-second clock effect—seems to take exactly this form: all of a sudden, the availability of clocks is assumed. Notably though, this specific issue need not be conceived as one of circularity: the *ad hoc* move can first of all be seen as a placeholder for a more detailed account on this specific transition. In fact, as we have already noted, Perlick (1987, 2008) has shown how the EPS scheme can be amended at this point by a constructivist clock construction in Weyl structures *prior* to assumptions about the second clock effect. The proponent of the EPS scheme has simply to concede that the original scheme was incomplete—but need not conclude that it was circular. A similar point can be made in the context of the objection of Sklar (1977) that the EPS approach is circular as it presupposes a distinction between forced versus force-free motion—again, if correct, the results of Coleman and Korté (1980) demonstrate simply that the approach is incomplete, rather than circular. (These additions to the EPS scheme will be assessed in more detail in Part II.)

---

[31]For example, on a (super-)Humean approach, one might begin with a Humean mosaic consisting of the motions of particles and light rays; construct therefrom via the EPS scheme the structure of a Lorentzian spacetime $(M, g_{\mu\nu})$; and then extract the dynamics as the simplest and strongest codification of these (themselves derivative) entities. For discussion of such (super-)Humean approaches, see e.g. (Pooley, 2013, §6.3.1).

The original EPS scheme also illustrates a more general stereotype regarding rational reconstructions: one must *fine-tune* the assumptions in the linear derivations in such a way that one gets what one is after, and nothing else. This fine-tuning then brings in a form of *methodological* circularity after all. The differentiability assumptions of EPS, for instance, do not follow from in-principle preconditions on measurement or modelling: they are at most what is pragmatically easiest to work with. Without further justification, they are problematic, as the final result of the rational reconstruction heavily depends on their invocation. Their only justification seems to lie in the fact that we *want* to derive the kinematic structure of general relativity.

Finally, the EPS scheme clearly violates mathematical linearity, i.e., the authors neither carry out—unsurprisingly, given the page constraints of single paper!—nor promote a constructivist mathematical program. It is not clear why linear constructivism when demanded in physics should not be demanded in mathematics as well. Compare EPS's silence here to to the stance taken by the school of 'proto-physics'—composed of Lorenzen, Janich and their students in Germany in the late 1960s to early 1980s—which aims at a non-circular, systematic account of how to set up basic measurement operations for those quantities that are essential to physical theorising—such as distance, time, and mass.[32] Their explanatory standard requiring a linear explanation for measurement operations likewise made them—arguably even as a precondition for their originally-planned undertaking—engage with a linear construction of logic, mathematics and language more generally.

## 8.2    Amendments

The EPS scheme has been the subject of various variations:

**Replacement of non-constructivist axioms**  As depicted before, non-linear reference to clocks can be remedied through the Weyl structure clock construction à la Perlick (1987). Again, as depicted before, Sklar's non-linearity objection has arguably been addressed by Coleman-Korté (see Coleman and Korté (1992), among others).

---

[32]Very crudely, its main motivation derives from a dissatisfaction with standard theory-based accounts of measurement, all of which suffer (the claim goes) from circularity in one form or another; its main strategy to evade such circularity lies in grounding measurement semantics in pragmatics. A collection edited by Böhme (1976) (in German only) and a monograph by Janich (2012) serve as good entry points to the literature.

**Alternative basic entities**  Woodhouse (1973) provides a linear set-up of manifold structure from causality assumptions not so different from that within the EPS scheme; this is an improvement on (parts of) the EPS scheme if any assumption of manifold is seen as to be reduced ultimately to causal relations (the EPS scheme assumes particles to be one-dimensional manifolds; cf. axiom $D_1$). The scheme of EPS has also been changed to be founded on empirically-motivated axioms regarding alternative probing structures, in particular to particles with spins and matter waves as test particles (see Audretsch and Lämmerzahl (1995) and Audretsch and Lämmerzahl (1991) respectively).

**Generalisations through loosening of assumptions**  Several assumptions (some of them made only implicitly) are not empirically motivated: (i) The implicit assumption of EPS that the manifold satisfies the Hausdorff condition cannot—at least at the point of introduction in the constructional system—be empirically motivated, and may thus be dropped altogether. Upon weakening this assumption, the question becomes whether at some later stage a suitable empirical axiom can be introduced that imposes the Hausdorff condition—or whether the EPS scheme should be widened into a constructive account of the kinematical space of non-Hausdorff GR (see Luc (2020); Luc and Placek (2020)). (ii) In axiom $P_2$, EPS appear to assume implicitly that coordinate systems are holonomic, thereby excluding torsionful spacetimes by fiat. Without so assuming that coordinate systems are holonomic, one opens the possibility that the axiomatisation will yield torsionful spacetimes (although, of course, it is also possible that by weakening axiom $P_2$ in this way, the entire ensuing structure of the axiomatisation is modified, thereby yielding further differences in the final class of spacetime kinematical possibilities at which one arrives[33]). (iii) The explicit assumption that $g$, as defined in the axiom, is a function of class $C^2$ on $U$ is (again) a strong idealisation. Upon weakening the assumption, the EPS scheme can be deformed into a constructive scheme for at least a subclass of Finslerian spacetimes (of which Lorentzian spacetimes are a special case): see Pfeifer (2019), section II C, and also Lämmerzahl and Perlick (2018).

**Rigorised proofs**  The central theorem on the existence of a Weyl metric has been given a rigorous proof by Matveev and Scholz (2020). (This is in so far a re-assuring result as the validity of the theorem had been put in doubt earlier by Matveev and Trautman (2014).)

## 8.3   Readings

In light of the above walkthrough to the EPS construction, we note the following specific readings:

**EPS as a what-is-sufficient explanation of kinematical structure:**  According to (Ehlers, 1973a, p. 22-23),

---

[33]Notably, the assumption of torsion-free affine connection is also used in showing the equivalence between constructive and algebraic notion of projective structure in the set-up; dropping it, might thus also be problematic at this point.

> [the] main purpose [of the EPS scheme] is to elucidate from
> a different point of view and in a systematic manner, which
> facts and mathematical idealizations of facts are sufficient
> to support the edifice of general relativity theory. In partic-
> ular the following approach enables one to isolate certain
> relatively independent substructures which are contained
> in Einstein's theory, the differential-topological, conformal,
> projective, affine and metric structures, which are related to
> different observable phenomena or to different relations in-
> ferred from such phenomena. These substructures are present
> also in theories different from Einstein's and might also be
> considered as building blocks for further developments needed
> when Einstein's theory has to be modified.

This arguably complements a what-empirical-structure-is-necessary
explanation of kinematical structure through the usual theoretical ac-
count which implies all of the used axioms by EPS.

**EPS as a how-possible explanation of theory change:** Although it is not
clear from the EPS scheme that the probing assumptions are meant
necessarily as probing assumptions from a local special relativistic
spacetime,[34] one might, in any case, argue that the EPS scheme can
be understood to provide a conceptualisation of how theory change
is possible from one theory to another by showing how the locali-
sation on probing assumptions leads to novel kinematical structure
(Lorentzian spacetime structure instead of Minkowski spacetime). Thus
read, the EPS scheme can be seen to provide a *how-possible* explana-
tion of how a shift from special to general relativistic kinematics could
have come about.[35]

**EPS as an ontological account of spacetime structure** The EPS program (at
least when freed of circularity) can be re-construed as ontologically re-
ducing spacetime structure to an ontology of particles and light rays.
Albeit prima facie just one out of several options to provide a fun-
damental ontology to GR, the ontological basis suggested by the EPS
scheme may be singled out by a view of naturalised metaphysics ac-
cording to which a fundamental ontology primarily is to be simple
and informative.[36]

Furthermore, there is the more intricate question of how to relate the
EPS scheme to past programs of linearly constructing spacetime structure
out of more primitive structure, such as (a) Klein's Erlanger program, (b) Re-
ichenbach's constructive axiomatisation, (c) Protophysics, and (d) Brown's
dynamical approach and other accounts of physical geometry. This will be
some of the major topics of Part II.

---

[34] At least, light is tacitly assumed to propagate with a finite propagation speed in the EPS scheme.

[35] Cf. the how-possible explanations of how general relativ kinematics can be motivated from
special relativistic kinematics in (Mittelstaedt, 2013, Ch. 2).

[36] Cf. Esfeld (2020) and other proponents of the minimalist ontology approach to quantum me-
chanics and physics more generally. See also our remarks on (super-)Humeanism in footnote 31.

## Acknowledgements

We are very grateful to Emily Adlam, Juliusz Doboszewski, Dennis Lehmkuhl, Niels Martens, Erhard Scholz, Chris Smeenk, Noah Stemeroff, to the attendees of the Bonn work in progress group, and to the Harvard Black Hole Institute foundations group, for valuable feedback. The diagrams in this article have been reproduced from those in the original article, the copyright for which is held by Oxford University Press and the Royal Irish Academy of Sciences; we are grateful to Andrzej Krasinski for providing us with guidance on the copyright issues related to these images.